\documentclass[%
 reprint,%
superscriptaddress,
 amssymb, amsmath,%
 aps,pre,%
]{revtex4-1}
\usepackage{dcolumn}
\usepackage{graphicx}
\usepackage{xcolor}
\usepackage{amsfonts}
\usepackage{amsmath}
\usepackage{amssymb}
\usepackage{amsthm}
\usepackage{bm}
\usepackage{hyperref}
\usepackage{comment}
\usepackage{subfigure}
\usepackage{mathtools}
\hypersetup{
  colorlinks   = true, %Colours links instead of ugly boxes
  urlcolor     = blue, %Colour for external hyperlinks
  linkcolor    = blue, %Colour of internal links
  citecolor   = blue %Colour of citations
}

\theoremstyle{remark}
\newtheorem{remark}{Remark}

\graphicspath{{figures/}}

\usepackage{moreverb}
\usepackage[colorinlistoftodos]{todonotes}

\newcounter{bgcomment}

\newcounter{gagcomment}

\newcounter{ldscomment}

\date{\today}

\begin{document}

\title{Model Reduction for the Kuramoto-Sakaguchi Model: The Importance of Non-entrained Rogue Oscillators}
%A Collective Coordinate Approach for the Kuramoto-Sakaguchi Model of Coupled Phase Oscillators

\author{Wenqi~Yue}
 \email{wyue8667@uni.sydney.edu.au}
 %\altaffiliation[Also at ]{Physics Department, XYZ University.}%Lines break automatically or can be forced with \\
   \affiliation{ 
School of Mathematics and Statistics, The University of Sydney, Sydney, New South Wales, 2006 Australia%\\This line break forced with \textbackslash\textbackslash
}%
\author{Lachlan~D. Smith}
\email{lachlan.smith@sydney.edu.au}
\affiliation{
	School of Mathematics and Statistics, The University of Sydney, Sydney, New South Wales, 2006 Australia}
\author{Georg~A. Gottwald}
\email{georg.gottwald@sydney.edu.au}
 \affiliation{ 
School of Mathematics and Statistics, The University of Sydney, Sydney, New South Wales, 2006 Australia%\\This line break forced with \textbackslash\textbackslash
}%

%%%%%%%%%%%%%%%%%%%%%%%%%%%%%%%%%%%%%%%%%%%

\begin{abstract}  
The Kuramoto-Sakaguchi model for coupled phase oscillators with phase-frustration is often studied in the thermodynamic limit of infinitely many oscillators. Here we extend a model reduction method based on collective coordinates to capture the collective dynamics of finite size Kuramoto-Sakaguchi models. We find that the inclusion of the effects of rogue oscillators is essential to obtain an accurate description, in contrast to the original Kuramoto model where we show that their effects can be ignored. We further introduce a more accurate ansatz function to describe the shape of synchronized oscillators. Our results from this extended collective coordinate approach reduce in the thermodynamic limit to the well-known mean-field consistency relations. For finite networks we show that our model reduction describes the collective behavior accurately, reproducing the order parameter, the mean frequency of the synchronized cluster, and the size of the cluster at given coupling strength, as well as the critical coupling strength for partial and for global synchronization. 
\end{abstract}

%%%%%%%%%%%%%%%%%%%%%%%%%%%%%%%%%%%%%%%%%%%

\maketitle

%%%%%%%%%%%%%%%%%%%%%%%%%%%%%%%%%%%%%%%%%%%

%%%%%%%%%%%%%%%%%%%%%%%%%%%%%%%%%%%%%%%%%%%

\section{Introduction} 
\label{sec:intro}

Synchronization is a ubiquitous phenomenon observed across a variety of different natural and artificial systems \cite{Kuramoto84,PikovskyEtAl01,Strogatz_2004}, from pace-maker cells of circadian rhythms \cite{Yamaguchi1408}, networks of neurons \cite{Bhowmik_Shanahan_2012} to chemical oscillators \cite{KissEtAl2002,TaylorEtAl2009} and power grid systems \cite{FilatrellaEtAl08}. A paradigmatic model for studying the dynamics of synchronization is the celebrated Kuramoto model of sinusoidally coupled phase oscillators \cite{Kuramoto84, Strogatz_2000, PikovskyEtAl01, AcebronEtAl05, OsipovEtAl07, ArenasEtAl08, DorflerBullo14, RodriguesEtAl16}. Real-world oscillatory systems are often prone to time-delayed or phase-frustrated coupling which are not described by the Kuramoto model. To capture the effects of phase-frustration, the Kuramoto model was extended to the Kuramoto-Sakaguchi model \cite{Sakaguchi-Kuramoto_1986}. Phase frustration, often associated with time-delayed couplings \cite{CrookEtAl97}, is important in various physical contexts including arrays of Josephson junctions \cite{WiesenfeldEtAl96,BarbaraEtAl99,FilatrellaEtAl00}, power grids \cite{NishikawaMotter15} and seismology \cite{Scholz10,VasudevanEtAl15}. Non-zero phase-frustration leads to a synchronized cluster rotating collectively with a non-zero frequency in the rest frame, in contrast with the Kuramoto Model, for which the synchronized cluster is stationary in the rest frame provided the intrinsic frequencies are symmetric about zero. Besides the familiar behavior of transitioning from incoherence through partial synchronization to full synchronization with increasing coupling strength, the Kuramoto-Sakaguchi model displays much richer dynamical behavior, such as bi-stability of incoherence and partial synchronization, transition from coherence to incoherence with increasing coupling strength \cite{Omelchenko2012, Omelchenko2013} as well as chaotic dynamics \cite{BickEtAl18}. Furthermore, the Kuramoto-Sakaguchi model has been the showground to study chimera states, where identical oscillators evolve to a state of coexistence between synchronization in some part and incoherence in other parts \cite{AbramsEtAl_2008, Laing_Chaos2009, MartensEtAl2016}. 

To understand and describe this plethora of collective dynamical scenarios, one seeks to derive reduced equations which facilitate analysis while still capturing the essential dynamics.  Previous model reduction methods for Kuramoto-like models primarily consider the thermodynamic limit of infinitely many oscillators. For the Kuramoto model without phase-frustration, the Ott-Antonsen (OA) ansatz achieves a reduction to a one-dimensional equation for the order parameter \cite{OttAntonsen08}. In the Kuramoto-Sakaguchi model a frequency-dependent version of the OA ansatz was developed to describe the non-zero rotation frequency of the synchronzied cluster \cite{Omelchenko2012,Omelchenko2013}. However, these model reduction methods only apply to the thermodynamic limit of infinitely many oscillators. Real world systems are of finite size and the behavior of finitely many oscillators may strongly deviate from their thermodynamic limit \cite{Gottwald_2017}. To relax the restriction of the thermodynamic limit used in mean-field theory and in model reduction approaches such as the OA ansatz, a model reduction framework using \textit{collective coordinates} has been developed recently for the original Kuramoto model \cite{Gottwald_2015}. The collective coordinate approach has been applied successfully to capture finite size effects in the original Kuramoto model \cite{Gottwald_2015} and in a stochastic Kuramoto model \cite{Gottwald_2017}, and has been extended to coupled oscillator models with arbitrary network topology \cite{HancockGottwald_2018}. For the Kuramoto model with multimodal intrinsic frequency distributions, the collective coordinate framework was used to describe collective chaos with multiple interacting synchronized clusters \cite{SmithGottwald_2018}. It has been established recently that in the thermodynamic limit, the collective coordinate framework recovers exact bifurcation structure of the OA ansatz \cite{SmithGottwald_2019}. 

In this paper we extend the approach to the Kuramoto-Sakaguchi model with non-zero phase-frustration. We show that to accurately reproduce the macroscopic dynamics, we must account for the non-entrained rogue oscillators, i.e., those that do not partake in the collective synchronized behavior. The rogue oscillators significantly affect the collective behavior of the synchronized cluster via their mean-field. The influence of the rogue oscillators on the synchronized oscillators is particularly prominent close to the onset of synchronization where the number of rogue oscillators is larger than the number of synchronized oscillators. The inclusion of the effect of the rogue oscillators requires careful analytical treatment. This is achieved by judiciously considering their average effect with respect to a specific probability distribution function. This is in contrast with the original Kuramoto model with no phase-frustration, for which we show that the effect of the rogue oscillators can be ignored for symmetric intrinsic frequency distributions. We apply the collective coordinate approach with two ansatz functions, a linear ansatz that corresponds to linearization of a mean-field solution and a fully non-linear arcsine ansatz. The arcsine ansatz has higher accuracy and recovers classical self-consistency results in the thermodynamic limit. The collective coordinate approach captures the transitions from incoherence to synchronized states. %
% , including the explosive first-order transition for a uniform intrinsic frequency distribution at zero phase-frustration, which is replaced by a more gradual, second order transition when the phase-frustration is non-zero. 
Our collective coordinate approach accurately captures several other collective quantities of the Kuramoto-Sakaguchi model, such as cluster mean frequency, cluster size and critical coupling strength corresponding to onset of partial and of global synchronization.

The paper is organized as follows. Section \ref{sec:KS_model} introduces the Kuramoto-Sakaguchi model and discusses some of its collective behavior. Section \ref{sec:mean_field} reviews the self-consistency analysis for the Kuramoto-Sakaguchi model in the thermodynamic limit.  Section \ref{sec:collective_coordinate} revisits the collective coordinate framework and develops its generalization for the Kuramoto-Sakaguchi model, in particular how to incorporate the dynamical effects of the rogue oscillators. Section \ref{sec:results} presents numerical results of the collective coordinate approach, showcasing its effectiveness to quantitatively describe the collective behavior of phase-frustrated oscillators. Section \ref{sec:conclusion} concludes with a summary and an outlook.

%%%%%%%%%%%%%%%%%%%%%%%%%%%%%%%%%%%%%%%%%%%

\section{The Kuramoto-Sakaguchi Model} \label{sec:KS_model}

The Kuramoto-Sakaguchi model describes the dynamics of $N$ coupled phase oscillators. It has the form
\begin{equation}{\label{KS-1}} 
\dot{\phi}_i(t)=\omega_i+\frac{K}{N}\sum_{j=1}^{N} \sin(\phi_j-\phi_i-\lambda),
\end{equation}
where $\phi_i(t)$ denotes the phase of the $i$-th oscillator with intrinsic frequency $\omega_i$, $K$ represents the strength of the coupling and $\lambda$ describes the phase-frustration. The constant phase lag is often viewed as an approximation for a time-delayed coupling when the delay is small \cite{CrookEtAl97}. The intrinsic frequencies $\omega_i$ are drawn from a frequency distribution $g(\omega)$. We consider here a Lorentzian distribution 
\begin{equation} \label{lorentzian_distribution}
g(\omega)=\frac{\Delta}{\pi(\Delta^2+\omega^2)} 
\end{equation}
with $\Delta=0.5$, and a uniform distribution
\begin{equation} \label{uniform_distribution}
g(\omega) \sim U[-\gamma,\gamma]
\end{equation}
with $\gamma=1$. The oscillators are ordered and indexed with increasing  intrinsic frequency $\omega_i$, i.e., $i=1$ corresponds to the smallest and $i=N$ to the largest intrinsic frequency. To mitigate against finite sampling effects such as frequency clusters we shall consider frequencies that are drawn from $g(\omega)$ equiprobably \footnote{To draw $N$ frequencies $\omega_i$ equiprobably from a density function $g(\omega)$, we let $x_i=\frac{1}{N}(i-\frac{1}{2})$ for $i=1,2\dots N$, let $f(\omega)=\int_{-\infty}^{\omega}g(s)ds$ be the cumulative density function of $g(\omega)$ and solve $f(\omega_i)=x_i$ for each $i$ to obtain the frequencies $\omega_i$.}. This avoids that a particular realization of the frequencies leads to either large or small gaps in the frequencies which implies local clustering. Finite size effects are still dynamically relevant as they determine the range of coupling strengths before new oscillators can be entrained, as will be discussed below. 

To describe the collective behavior of the Kuramoto-Sakaguchi model, mean-field variables $r$ and $\psi$ are introduced such that
\begin{align}{\label{mean_field}} 
re^{i \psi}=\frac{1}{N} \sum_{j=1}^{N} e^{i \phi_j}.
\end{align}
The time average of the order parameter $r$,
\[\bar{r}=\lim\limits_{T \rightarrow \infty}\frac{1}{T}\int_{T_0}^{T_0+T}r(t)dt\] quantifies the degree of synchronization. Synchronized states correspond to $\bar{r} \approx 1$, whereas incoherent states correspond to $\bar{r} \sim 1/\sqrt{N}$.

In the following we present results on the collective behavior of the Kuramoto-Sakaguchi model (\ref{KS-1}) which will be captured quantitatively via the collective coordinate approach developed in Section~\ref{sec:collective_coordinate}.\\

\begin{figure}[tbp]
	\centering
	\includegraphics[width=1\linewidth]{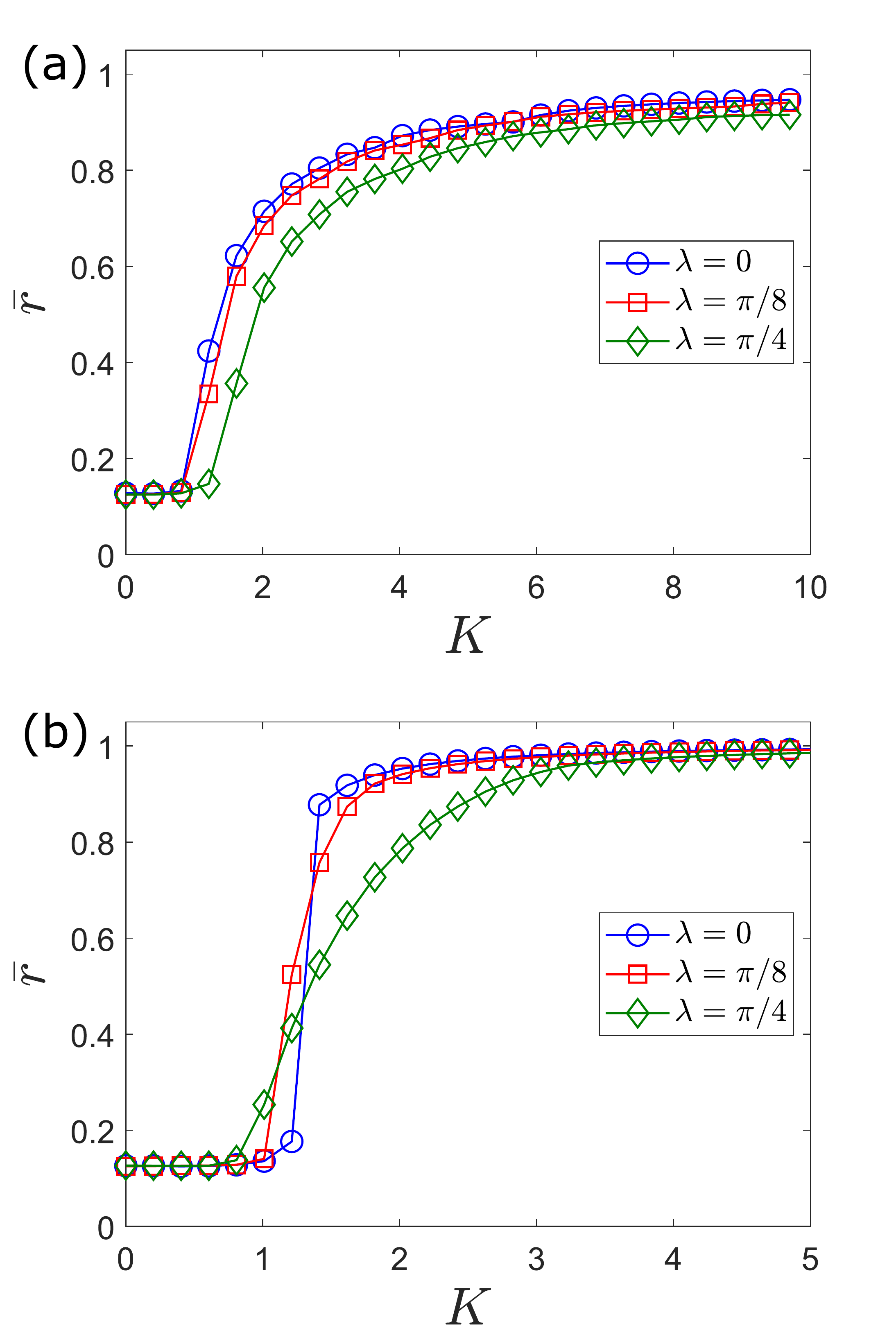}
	\caption{Order parameter $\bar{r}$ of the Kuramoto-Sakaguchi model (\ref{KS-1}) with $N=50$ oscillators for different phase-frustration parameters $\lambda$. (a) Lorentzian frequency distribution (\ref{lorentzian_distribution}). (b) Uniform frequency distribution (\ref{uniform_distribution}).}
	\label{fig:rbar_vs_K}
\end{figure}

Fig.~\ref{fig:rbar_vs_K} depicts the transition from incoherence with $\bar{r}\sim 1/\sqrt{N}$ to synchronization with $\bar{r}>0$ upon increasing the coupling strength $K$ beyond a critical coupling strength $K_c$ for (a) the Lorentzian frequency distribution (\ref{lorentzian_distribution}) and (b) the uniform frequency distribution (\ref{uniform_distribution}). For a Lorentzian distribution, inclusion of the phase-frustration $\lambda$ impedes synchronization, both lowering $\bar{r}$ for a given coupling strength $K$ as well as delaying the onset of synchronization to a higher value of $K$ for larger $\lambda>0$. For $K>K_c$ a partially synchronized cluster emerges, which increases in size upon increasing the coupling strength. Since a Lorentzian frequency distribution has an unbounded support in the thermodynamic limit $N\rightarrow\infty$, for each value of $K$ there are oscillators which are not entrained and hence global synchronization, in which all oscillators partake in the synchronized collective behavior, does not occur in the thermodynamic limit. For finite systems, however, there always exists a coupling strength $K_g$ above which all oscillators are synchronized. We remark that for $N=50$ oscillators at $\lambda=\pi/4$ with intrinsic frequencies drawn equiprobably from a Lorentzian distribution with $\Delta=0.5$, as shown in Fig.~\ref{fig:kc_comparison}(a), the transition to global synchronization occurs at $K_g\approx53.7$, which is outside the range of the figure. 

For a uniform distribution, increasing $\lambda$ similarly increases the critical coupling strength corresponding to global synchronization (not shown). However, unlike for a Lorentzian distribution, the onset of partial synchronization around $K\approx 1$ occurs at lower values of $K$ as $\lambda$ increases. For $\lambda=0$, it is well known that the transition is explosive from incoherence to global synchronization \cite{Pazo05}.\\ 
%In addition, it is observed that the well-known explosive, first-order transition from incoherence to global synchronization for the Kuramoto model with $\lambda=0$ \cite{Pazo05} becomes a second-order transition in the Kuramoto-Sakaguchi model with non-zero $\lambda$.\\
 
For $K>K_c$, a subset of the oscillators form a synchronized cluster which collectively evolves at a common non-zero frequency. To identify those oscillators that partake in the synchronized cluster, we compute the effective frequency $\Omega_i=\langle\dot{\phi}_i(t)\rangle_t$ for each oscillator where $\dot{\phi}_i(t)$ are instantaneous frequencies and $\langle - \rangle_t$ denotes a temporal average. Oscillators with a common $\Omega_i$ are identified as the synchronized cluster $\mathcal{C}$, with  indices $i_{\text{min}}\leq i \leq i_{\text{max}}$, and minimal and maximal intrinsic frequencies $\omega_{\text{min}}:=\omega_{i_{\text{min}}}$ and $\omega_{\text{max}}:=\omega_{i_{\text{max}}}$. The common frequency of the cluster $\mathcal{C}$ is estimated as $\Omega=\frac{1}{N_c}\sum_{j\in \mathcal{C}}\Omega_j$, where $N_c=|\mathcal{C}|=i_{\text{max}}-i_{\text{min}}+1$ denotes the size of the synchronized cluster.

\begin{figure}[tbp] 
	\includegraphics[width=1\linewidth]{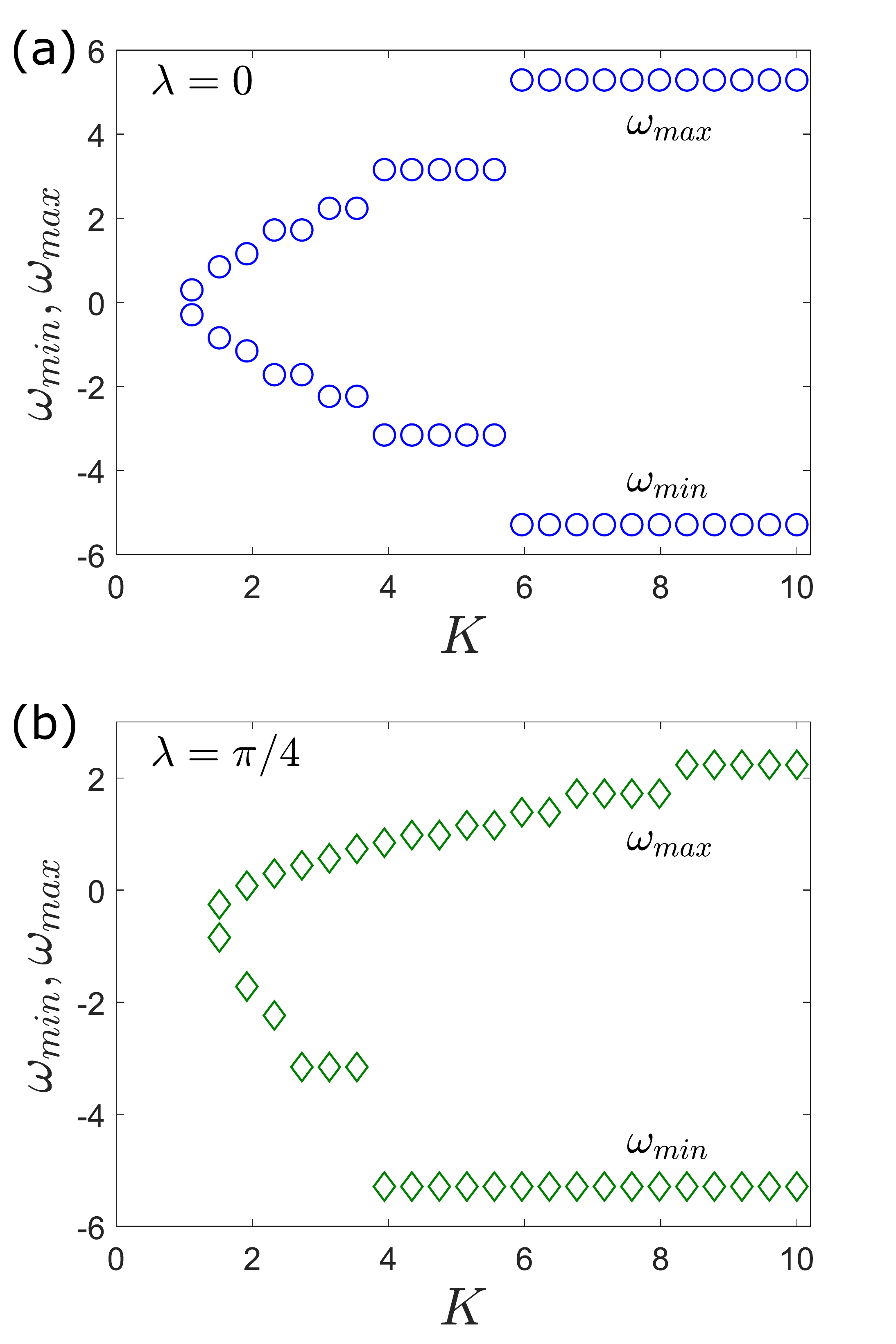}
	\caption{Minimum ($\omega_{\text{min}}$) and maximum ($\omega_{\text{max}}$) intrinsic frequencies of oscillators within the synchronized cluster at different coupling strengths $K$ for $N=50$ oscillators. Intrinsic frequencies drawn from a Lorentzian distribution (\ref{lorentzian_distribution}). (a) $\lambda=0$, (b) $\lambda=\pi/4$.}  
	\label{fig:position_of_cluster}
\end{figure}

In the original Kuramoto model with zero phase-frustration $\lambda=0$, synchronized clusters of size $N_c$ are always symmetric about $\omega=0$ provided the intrinsic frequencies $\omega_i$ are symmetric about $\omega=0$ (as is the case for equiprobable draws of a frequency distribution $g(\omega)$ that is symmetric about $\omega=0$). In particular for $\lambda=0$ we have $\omega_{\text{max}}=-\omega_{ \text{min}}$ and as the coupling strength $K$ decreases, oscillators break off from the cluster in symmetric pairs as shown in Fig.~\ref{fig:position_of_cluster}(a). In contrast, in the Kuramoto-Sakaguchi model with non-zero phase-frustration $\lambda$, the synchronized cluster is not symmetric about $\omega=0$ and oscillators break off from the cluster asymmetrically, as shown in Fig.~\ref{fig:position_of_cluster}(b) for $\lambda=\pi/4$.
 
\begin{figure}[tbp] 
	\centering 
	\includegraphics[width=1\linewidth]{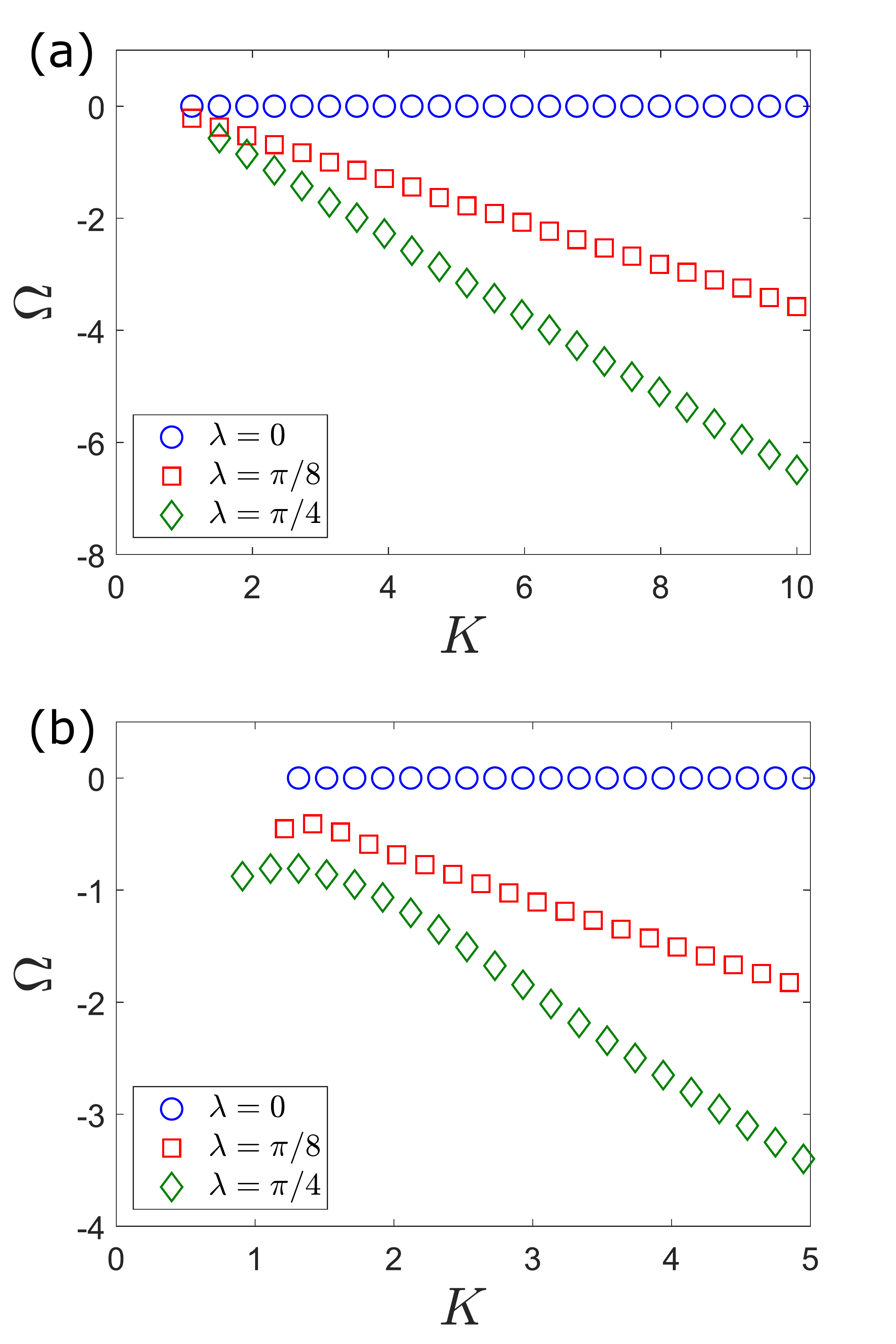}	 	 
	\caption{Mean frequency $\Omega$ of the synchronized cluster as a function of coupling strength $K$ for different phase-frustration parameters $\lambda$ for the Kuramoto-Sakaguchi model (\ref{KS-1}) with $N=50$ oscillators. (a) Lorentzian frequency distribution (\ref{lorentzian_distribution}). (b) Uniform frequency distribution (\ref{uniform_distribution}).
	} 
	\label{fig:Omega_vs_K}
\end{figure}

For non-zero $\lambda$, the synchronized cluster rotates at a non-zero common frequency $\Omega$ in the rest frame, in contrast to the Kuramoto model at $\lambda=0$ where the synchronized cluster is stationary. Fig.~\ref{fig:Omega_vs_K} shows the cluster mean frequency $\Omega$ as a function of the coupling strength $K$ for different values of $\lambda$. We observe that at high coupling strength, $\Omega$ has a nearly linear dependence on $K$ with $\Omega\approx-K\sin\lambda$ for both a Lorentzian frequency distribution (\ref{lorentzian_distribution}) and a uniform frequency distribution (\ref{uniform_distribution}). For the Lorentzian distribution, the linear dependence extends over the whole range of coupling strengths. For the uniform distribution, $\Omega$ exhibits a non-linear dependency of $\Omega(K)$ close to the onset of partial synchronization.

In the following we present the well-known self-consistency result obtained from mean-field analysis in the thermodynamic limit. These results will be used subsequently in our model reduction via the collective coordinate approach, and will be instructive in extending the collective coordinate approach developed in \cite{Gottwald_2015,Gottwald_2017,HancockGottwald_2018,SmithGottwald_2018,SmithGottwald_2019} to incorporate rogue oscillators.

%%%%%%%%%%%%%%%%%%%%%%%%%%%%%%%%%%%%%%%%%%%

\subsection{Classical Mean-Field Approach and Self-Consistency Analysis} 
\label{sec:mean_field}

The Kuramoto-Sakaguchi model allows for a mean-field description in the thermodynamic limit, which reduces the dynamics to two mean-field variables $r$ and $\Omega$. Here we follow the approach developed by Sakaguchi and Kuramoto \cite{Sakaguchi-Kuramoto_1986} to find a self-consistency relation between $r$ and $\Omega$. We shift to the frame rotating with the cluster mean frequency $\Omega=\Omega(K)$ and consider the phase variables $\theta_i(t)=\phi_i(t)-\Omega t$. The Kuramoto-Sakaguchi model (\ref{KS-1}) is then written as
\begin{equation} \label{KS-theta}
\dot{\theta}_i(t)=\omega_i-\Omega+\frac{K}{N}\sum_{j=1}^{N} \sin(\theta_j-\theta_i-\lambda),
\end{equation}
and the mean-field variables (\ref{mean_field}) are expressed as
\begin{equation}{\label{mean_field_rotating_frame}} 
re^{i \psi}=\left(\frac{1}{N}\sum_{j=1}^{N}e^{i \theta_j}\right)e^{i \Omega t}.
\end{equation}

In the thermodynamic limit, after a sufficiently long transient, $r$ and $\Omega$ asymptote toward steady states, and without loss of generality we set $\psi=\Omega t$. Substituting (\ref{mean_field_rotating_frame}) into the Kuramoto-Sakaguchi model (\ref{KS-theta}) we obtain the mean-field formulation of the Kuramoto-Sakaguchi model,
\begin{equation} {\label{KS-mean-field}}
\dot{\theta}_i(t)=\omega_i-\Omega-Kr\sin(\theta_i+\lambda).
\end{equation}
In this form, $\theta_i$ is coupled to the phases of the other oscillators only via the mean-field variables $r$ and $\Omega$. 

Oscillators with frequencies $|\omega_i-\Omega|\leq Kr$ allow for stationary solutions
\begin{equation} \label{fixed_point_synchronized_cluster}
\theta_i=\arcsin\left(\frac{\omega_i-\Omega}{Kr}\right)-\lambda,
\end{equation}
and partake in the collective rotation with frequency $\Omega$. These oscillators form the synchronized cluster $\mathcal{C}$. On the other hand, oscillators with frequencies $|\omega_i-\Omega|>Kr$ do not allow for fixed point solutions and instead drift with
\begin{equation*}
\dot{\theta}_i=v(\theta_i;\omega_i),
\end{equation*}
where
\begin{equation} \label{v_mean_field}
v(\theta_i;\omega_i)=\omega_i-\Omega-Kr\sin(\theta_i+\lambda).
\end{equation}
These oscillators are the non-entrained, rogue oscillators.\\

In the thermodynamic limit $N \rightarrow \infty$, the phases can be described by a probability density function $\rho(\theta,t;\omega)$ with normalization condition
\begin{equation}
\label{eq:norm} 
\int_0^{2\pi}\rho(\theta,t;\omega)d\theta=1.
\end{equation}
The probability density $\rho(\theta,t;\omega)$ satisfies the continuity equation
\begin{equation} \label{continuity_eqn_non_entrained}
\frac{\partial \rho}{\partial t}+\frac{\partial}{\partial \theta}(\rho v)=0,
\end{equation}
which is a consequence of conservation of the number of oscillators at each intrinsic frequency. We now consider particular stationary solutions of the continuity equation (\ref{continuity_eqn_non_entrained}), pertaining to the entrained synchronized oscillators and the non-entrained rogue oscillators.

After a sufficiently long transient, oscillators with $|\omega-\Omega|\leq Kr$ become entrained and are described by a stationary probability density function
\begin{equation} \label{rho_entrained}
\rho(\theta;\omega)=\delta(\theta-\arcsin\left(\frac{\omega-\Omega}{Kr}\right)+\lambda). 
\end{equation}
The non-entrained, rogue oscillators with $|\omega-\Omega|>Kr$ have the stationary phase distribution 
\begin{equation} \label{rho_rogue}
\rho(\theta;\omega)=\frac{C(\omega)}{v(\theta;\omega)}=\frac{C(\omega)}{\omega-\Omega-Kr\sin(\theta+\lambda)},
\end{equation}
where $C(\omega)$ is a normalization constant.

The mean-field variable (\ref{mean_field_rotating_frame}) can then in the thermodynamic limit be expressed as
\begin{align} \label{mean_field_thermodynamic_limit}
r=\int_{-\infty}^{\infty} \int_{0}^{2\pi}e^{i \theta}\rho(\theta,t;\omega) g(\omega) d\theta d\omega.
\end{align}
Substituting  the stationary probability densities for the entrained and for the non-entrained oscillators, (\ref{rho_entrained}) and (\ref{rho_rogue}), into the equation for the mean-field order parameter  (\ref{mean_field_thermodynamic_limit}), we obtain
\begin{align}
re^{i \lambda}&= \displaystyle\int\displaylimits_{\mathclap{|\omega-\Omega|\leq Kr}} \left(\sqrt{1-\frac{(\omega-\Omega)^2}{K^2r^2}}+ i\frac{\omega-\Omega}{Kr}\right) g(\omega)d \omega \nonumber\\
&+i\displaystyle\int\limits_{\mathclap{|\omega-\Omega|>Kr}}\left(\frac{\omega-\Omega}{Kr}-\frac{\omega-\Omega}{Kr}\sqrt{1-\frac{K^2r^2}{(\omega-\Omega)^2}}\right)g(\omega)d \omega. \label{self_consistency_1}
\end{align}
Considering real and imaginary parts of (\ref{self_consistency_1}), we arrive at the following self-consistency relation for $r$ and $\Omega$,
\begin{align} 
r \cos\lambda&=\displaystyle\int\displaylimits_{\mathclap{|\omega-\Omega|\leq Kr}} \sqrt{1-\frac{(\omega-\Omega)^2}{K^2r^2}} g(\omega) d\omega\label{self_consistency_real}\\
r \sin \lambda&= \displaystyle\int\displaylimits_{\mathclap{|\omega-\Omega|\leq Kr}}\frac{\omega-\Omega}{Kr}g(\omega)d\omega \nonumber\\
&+\displaystyle\int\limits_{\mathclap{|\omega-\Omega|>Kr}}\left(\frac{\omega-\Omega}{Kr}-\frac{\omega-\Omega}{Kr}\sqrt{1-\frac{K^2r^2}{(\omega-\Omega)^2}}\right)g(\omega)d \omega.
\label{self_consistency_imag}
\end{align}
Albeit complicated, in principle, for a specified distribution $g(\omega)$, the self-consistency equations (\ref{self_consistency_real})--(\ref{self_consistency_imag}) implicitly determine $r$ and $\Omega$ for a given coupling strength $K$ and phase-frustration $\lambda$.

\begin{remark} 
	For the Kuramoto model with $\lambda=0$, the left-hand side of (\ref{self_consistency_imag}) is zero. If $g(\omega)$ is symmetric about $\omega=0$, then (\ref{self_consistency_imag}) is satisfied for $\Omega=0$ and (\ref{self_consistency_real}) becomes 
	\begin{align} 
	    r=\int_{-Kr}^{Kr} \sqrt{1-\frac{\omega^2}{K^2r^2}}g(\omega) d \omega,
	\end{align}
	which is the classical self-consistency result for the Kuramoto model \cite{Kuramoto84,Strogatz_2000}.
\end{remark}

\begin{remark} 
If the intrinsic frequency distribution $g(\omega)$ has finite support then for large enough $K$, the synchronized cluster $|\omega-\Omega|\leq Kr$ contains all the oscillators and if additionally $g(\omega)$ is symmetric about $\omega=0$, (\ref{self_consistency_imag}) becomes
\begin{align} 
r\sin \lambda =& \int_{-\infty}^{\infty} \frac{\omega-\Omega}{Kr}g(\omega)d\omega +0 =-\frac{\Omega}{Kr},
\label{e.OmKr}
\end{align}
leading to $\Omega=-Kr^2\sin\lambda$. For high $K$, $r\approx 1$ and $\Omega\approx-K\sin\lambda$, agreeing with the linear dependence observed in Fig.~\ref{fig:Omega_vs_K}.
\end{remark}

In Appendix~\ref{sec:thermodynamics_1} we apply the Ott-Antonsen ansatz for a Lorentzian intrinsic frequency distribution to obtain expressions for $r(K,\lambda)$ and $\Omega(K,\lambda)$. We find that for the Lorentzian frequency distribution the transition to partial synchronization is via a supercritical pitchfork bifurcation. %For the uniform frequency distribution we find that for $\lambda\neq 0$ the transition is a 

%In Appendix~\ref{sec:thermodynamics_1} we evaluate the integrals in the self-consistency equations (\ref{self_consistency_real})--(\ref{self_consistency_imag}) for a Lorentzian intrinsic frequency distribution. We find that for the Lorentzian frequency distribution the transition to partial synchronization is via a supercritical pitchfork bifurcation. %For the uniform frequency distribution we find that for $\lambda\neq 0$ the transition is a smooth second order phase transition from incoherence to partial synchronization, whereas for $\lambda=0$ the transition is a discontinuous explosive first order phase transition straight to global synchronization, meaning that for $\lambda=0$ either all oscillators are synchronized or none.

%In Appendix~\ref{sec:thermodynamics_1} we evaluate the integrals in the self-consistency equations (\ref{self_consistency_real})--(\ref{self_consistency_imag}) for Lorentzian and for uniform intrinsic frequency distributions. We find that for the Lorentzian frequency distribution the transition to partial synchronization is via a supercritical pitchfork bifurcation. For the uniform frequency distribution we find that for $\lambda\neq 0$ the transition is a smooth second order phase transition from incoherence to partial synchronization, whereas for $\lambda=0$ the transition is a discontinuous explosive first order phase transition straight to global synchronization, meaning that for $\lambda=0$ either all oscillators are synchronized or none.

%%%%%%%%%%%%%%%%%%%%%%%%%%%%%%%%%%%%%%%%%%%

\section{Collective Coordinate Approach} 
\label{sec:collective_coordinate}

A model reduction method based on collective coordinates has recently been proposed and developed for finite-size Kuramoto models \cite{Gottwald_2015, Gottwald_2017, HancockGottwald_2018, SmithGottwald_2018, SmithGottwald_2019}. The approach is based on projecting the dynamics of the full model onto a judiciously chosen low-dimensional ansatz manifold to capture the collective dynamics of the system. The variables that are used to parameterize the ansatz manifold are the so-called collective coordinates.

For the Kuramoto-Sakaguchi model, at sufficiently high coupling strengths, a group of oscillators form a synchronized cluster with a well-defined shape profile that rotates at a constant frequency. Motivated by this observation, we propose that in the frame co-rotating at the cluster mean frequency $\Omega$, the phases of oscillators within the synchronized cluster $i \in \mathcal{C}$ are approximated by 
\begin{align} 
\theta_i(t)\approx \Theta_i(\alpha(t),\Omega),
\label{eq:cc}
\end{align}
where $\Theta_i$ is the shape profile of phases of oscillators within the cluster $\mathcal{C}$, $\alpha(t)$ is a collective coordinate representing the spread of the phases and $\Omega$ is a collective parameter representing the cluster mean frequency. Here the synchronized cluster $\mathcal{C}$ is specified \textit{a priori}, we will discuss later how $\mathcal{C}$ is determined via the collective coordinate approach. Note that $\Omega$ does not have explicit time dependence. The method of collective coordinates requires to specify the shape profile $\Theta$, and to determine the evolution of the collective coordinate $\alpha(t)$ and an expression for the rotation frequency $\Omega$. We begin by specifying ansatz functions for the shape $\Theta_i$ for the Kuramoto-Sakaguchi model (\ref{KS-theta}). We will specify two ansatz functions; a linear function, approximating the shape for large coupling strengths $K$, and a nonlinear function which describes the mean-field in the thermodynamic limit.

Linearization of the stationary mean-field solution (\ref{fixed_point_synchronized_cluster}) around $1/K\ll 1$ suggests a shape profile of the form 
\begin{equation} {\label{ansatz_linear}}
\Theta_i=\frac{\omega_i-\Omega}{Kr}-\lambda,
\end{equation}
by expanding  the stationary solution (\ref{fixed_point_synchronized_cluster}) up to linear order. We coin this the linear ansatz.

Alternatively, using the full nonlinear stationary mean-field solution (\ref{fixed_point_synchronized_cluster}), we arrive at 
\begin{equation} {\label{ansatz_arcsine}} 
\Theta_i=\arcsin\left(\frac{\omega_i-\Omega}{Kr}\right)-\lambda.
\end{equation}
We coin this the arcsine ansatz. Note that the arcsine ansatz is exact for globally synchronized systems. 

\begin{figure}[tbh] 
	\centering
	\includegraphics[width=1\columnwidth]{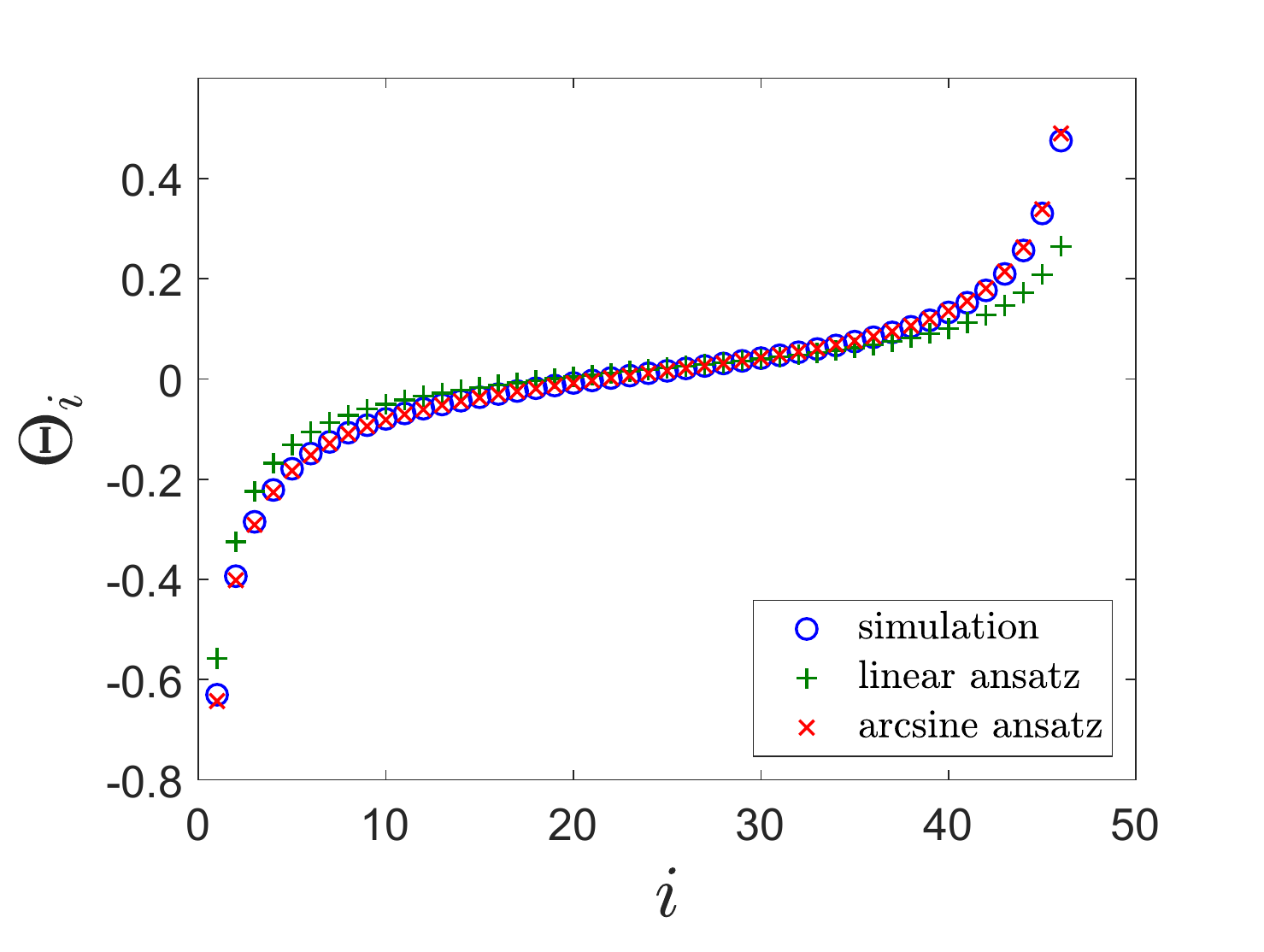}
	\caption{Stationary phases from the linear ansatz (\ref{ansatz_linear}), the arcsine ansatz (\ref{ansatz_arcsine}) and snapshot of the phases obtained from a simulation of the full Kuramoto-Sakaguchi model (\ref{KS-theta}) with $N=50$ oscillators at coupling strength $K=10$ and with phase-frustration parameter $\lambda=\pi/4$, and a Lorentzian frequency distribution (\ref{lorentzian_distribution}). The synchronized cluster forms for $2\leq i \leq 47$. Each set of phases is shifted to have mean zero.} 
	\label{fig:phases_compare} 
\end{figure}

For both ansatzes (\ref{ansatz_linear}) and (\ref{ansatz_arcsine}), we identify $\alpha(t)=r(t)$ as the collective coordinate, and $\Omega$ as the collective parameter that does not have explicit time dependence. Fig.~\ref{fig:phases_compare} compares the phases predicted by the linear ansatz (\ref{ansatz_linear}) and the arcsine ansatz (\ref{ansatz_arcsine}) with the phases obtained from simulations of the full Kuramoto-Sakaguchi model (\ref{KS-theta}). Here the values of $r$ and $\Omega$ which appear in the ansatzes (\ref{ansatz_linear}) and (\ref{ansatz_arcsine}) are taken from time-averages from the numerical simulation. We will see below how to obtain the order parameter and the frequency from the collective coordinate approach. The arcsine ansatz produces phases which are barely distinguishable with the naked eye from those obtained from the simulations. The linear ansatz is also found to approximate the actual phases reasonably well. 

\begin{figure}[tbh] 
	\centering
	\includegraphics[width=1\columnwidth]{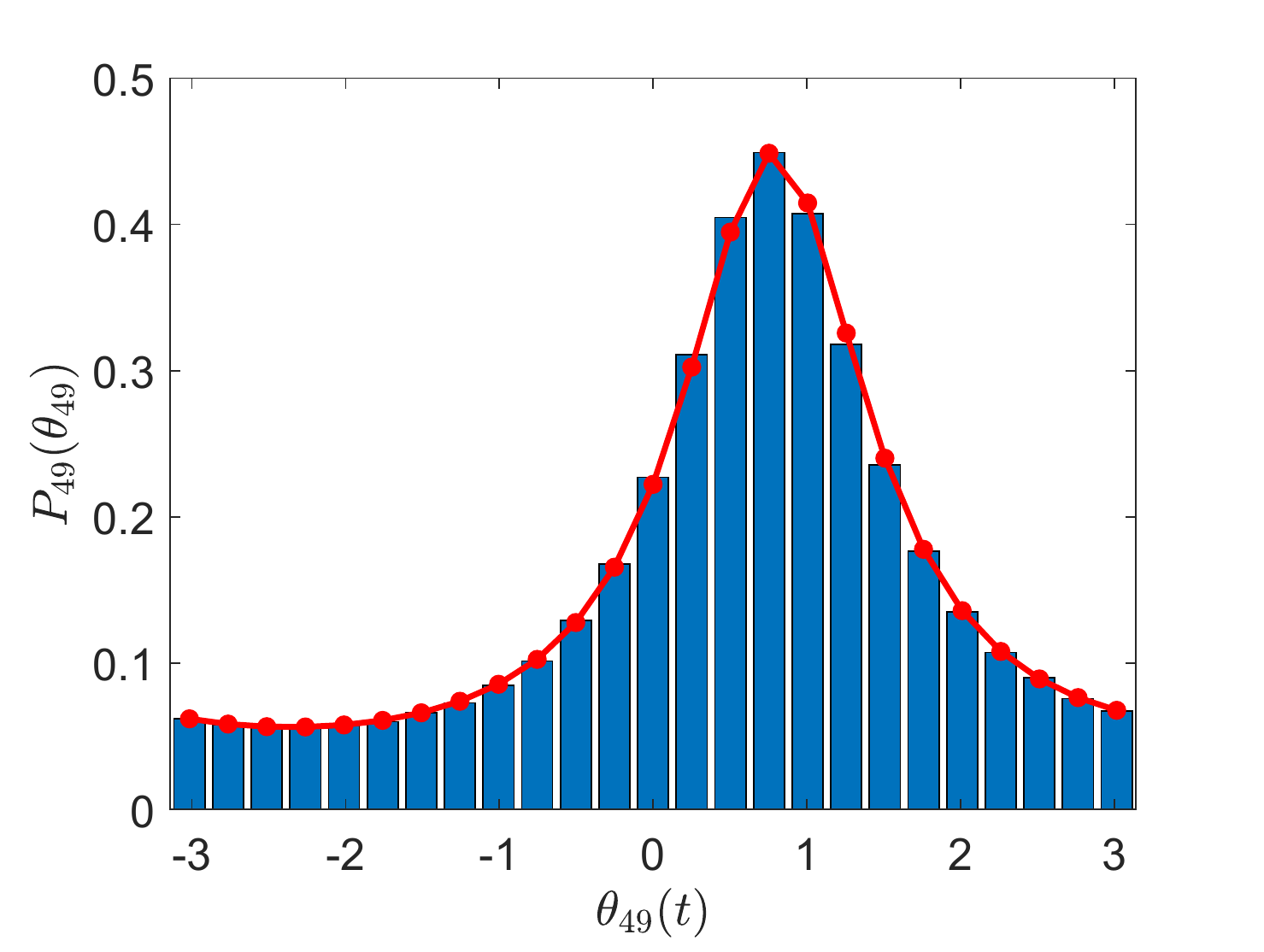}
	\caption{Normalized phase distribution of a single rogue oscillator $\theta_{49}$ obtained from numerical simulation of the full Kuramoto-Sakaguchi model  (\ref{KS-1}) (histogram) and obtained from the ansatz (\ref{ansatz_rogue_distribution}) for the density function $P_i(\theta_i)$  (continuous curve). Parameters are as in Fig.~\ref{fig:phases_compare}. Phases from simulation $\phi_i(t)$ are shifted to obtain the corresponding phases $\theta_i(t)$ in the reference frame rotating with frequency $\Omega$.}
	\label{fig:histagram_compare} 
\end{figure}

In previous work on collective coordinates model reduction \cite{Gottwald_2015, Gottwald_2017, HancockGottwald_2018, SmithGottwald_2018, SmithGottwald_2019}, non-entrained rogue oscillators $i \notin \mathcal{C}$ and their effect on the entrained oscillators, which are described by (\ref{eq:cc}), were ignored. Here we extend the collective coordinate framework to include the effect of non-entrained rogue oscillators, which will be shown in Section~\ref{sec:results} to be crucial to accurately reproduce the collective dynamics of the Kuramoto-Sakaguchi model. We propose that their phases are described by a distribution function inversely proportional to their instantaneous ``velocity''
\begin{align} \label{ansatz_rogue_distribution}
\theta_i \propto P_i(\theta_i)=\frac{C(\omega_i)}{v(\theta_i;\omega_i)},
\end{align}
where the velocity $v(\theta_i;\omega_i)$ is given by (\ref{v_mean_field}). Here $r$ and $\Omega$ appearing in the velocity (\ref{v_mean_field}) are given by the collective coordinates. This statistical ansatz for the rogue oscillators is motivated by the fact that $dt/d\theta_i = 1/v_i(\theta_i)$ measures the time the phase of the $i$-th oscillator near the value $\theta_i$, and, since the dynamics of the rogues is fast relative to the slow synchronization dynamics, the effect of the rogues on the entrained oscillators is determined by their statistical time-average. Fig.~\ref{fig:histagram_compare} provides numerical evidence for the statistical ansatz (\ref{ansatz_rogue_distribution}). We show a comparison of 
the proposed distribution function (\ref{ansatz_rogue_distribution}) with the normalized phase histogram for a single rogue oscillator $\theta_{49}$ obtained from a long-time simulation of the full Kuramoto-Sakaguchi model (\ref{KS-theta}).

We now set out to determine the evolution equations for the collective coordinates $\alpha(t)=r(t)$ and $\Omega$, where for the first time we incorporate the effect of the rogue oscillators. Following the collective coordinate framework \cite{Gottwald_2015, Gottwald_2017, HancockGottwald_2018, SmithGottwald_2018, SmithGottwald_2019} the dynamics of the collective coordinates is obtained by minimizing the error associated with the ansatz. The ansatz $\Theta_i$ (e.g. the linear ansatz (\ref{ansatz_linear}) or the arcsine ansatz (\ref{ansatz_arcsine})) is substituted into the original Kuramoto-Sakaguchi model (\ref{KS-theta}) to obtain the associated error for $i \in \mathcal{C}$,
\begin{align} {\label{associated_error}}
\mathcal{E}_i= & \dot{r}\frac{\partial \Theta_i}{\partial r}(r(t),\Omega)-(\omega_i-\Omega)\nonumber\\
&-\frac{K}{N}\left[\sum_{j \in \mathcal{C}}\sin(\Theta_j-\Theta_i-\lambda)+\sum_{j \notin \mathcal{C}}\sin(\theta_j-\Theta_i-\lambda)\right],
\end{align}
where we split the interaction term of entrained oscillators into a contribution coming from interactions with other entrained oscillators and a contribution coming from interactions with non-entrained rogue oscillators. 

The non-entrained rogue oscillators evolve on a time-scale much faster than the synchronized cluster, which is stationary in the co-rotating frame. This separation of time scales suggests that a synchronized node $\theta_i$, $i\in\mathcal{C}$ feels the time-averaged dynamics of the rogue oscillators $\theta_j$, $j\notin\mathcal{C}$. Invoking Birkhoff's ergodic theorem the temporal average can be approximated by averaging over the phase distribution $P_i(\theta_i)$ (\ref{ansatz_rogue_distribution}) and the contribution of the interaction term involving the rogue oscillators can be written as 
\begin{align*} 
\sum_{j \notin \mathcal{C}}\sin(\theta_j-\Theta_i-\lambda)&\approx \sum_{j \notin \mathcal{C}}\int_{0}^{2\pi}\sin(\theta_j-\Theta_i-\lambda)P_j(\theta_j)d\theta_j\\&=\cos(\Theta_i+2\lambda)\sum_{j\notin \mathcal{C}}k_j,
\end{align*}
where
\begin{equation} \label{k_j}
k_j=\frac{\omega_j-\Omega}{Kr}\left(1-\sqrt{1-\frac{K^2r^2}{(\omega_j-\Omega)^2}}\right).
\end{equation}
Note that for the original Kuramoto model with $\lambda=0$ and with intrinsic frequencies $\omega_i$ which are symmetric about $\omega=0$, the collective frequency $\Omega$ of the synchronized cluster is zero, which implies $\sum_{j\notin \mathcal{C}}k_j=0$. This justifies that for the Kuramoto model and symmetric frequency distributions, the rogue oscillators can be neglected, as was assumed in \cite{Gottwald_2015,Gottwald_2017,HancockGottwald_2018,SmithGottwald_2018,SmithGottwald_2019}. For a general Kuramoto-Sakaguchi model with $\lambda\neq0$, $\sum_{j\notin \mathcal{C}}k_j\neq 0$ in general. 

We shall see in the numerical simulations presented in Sec.~\ref{sec:results} that the inclusion of the rogue oscillators via the averaged effect on the synchronized cluster is crucial to obtain qualitative agreement between the reduced dynamics and the full Kuramoto-Sakaguchi model. Ignoring the interaction term by setting $\sum_{j\notin \mathcal{C}}k_j=0$ will be shown to only describe the collective behavior for large values of the coupling strength $K$ beyond the onset of global synchronization.\\

To obtain the dynamics of the collective coordinate $r(t)$, the error (\ref{associated_error}) is minimized, which is achieved when it is orthogonal to the tangent space of the ansatz manifold spanned by the collective coordinates $r$ and $\Omega$, i.e., when
\begin{align} 
\sum_{i \in \mathcal{C}}\mathcal{E}_i\frac{\partial \Theta_i}{\partial r}=0 \hspace{10pt} \text{and} \hspace{10pt} \sum_{i \in \mathcal{C}}\mathcal{E}_i\frac{\partial \Theta_i}{\partial \Omega}=0.
\label{eq:errormin}
\end{align} 
This yields a system of algebro-differential equations for the collective coordinates $r$ and $\Omega$ of the form 
\begin{align} 
\dot{r}&=F_a(r,\Omega) \label{reduced_equation_a}\\
0&=F_b(r,\Omega). \label{reduced_equation_b}
\end{align}
We provide explicit expressions for $F_{a,b}$ in Appendix~\ref{sec:reduced_equation} for the linear ansatz (\ref{ansatz_linear}) and the arcsine ansatz (\ref{ansatz_arcsine}).  We remark that solving the algebraic equation (\ref{reduced_equation_b}) yields $\Omega=\Omega(r)$, which, upon substitution into (\ref{reduced_equation_a}), gives an explicit evolution equation for the order parameter
\begin{equation} \label{reduced_equation_c}
\dot{r}=F_a(r,\Omega(r)).
\end{equation}

We are concerned with stationary solutions $\dot r = 0$, since stable stationary solutions correspond to synchronized states. For the arcsine ansatz with rogues included, the equations for stationary solutions of (\ref{reduced_equation_a})--(\ref{reduced_equation_b}) reduce to
\begin{align} 
r\cos\lambda&=\frac{1}{N}\sum_{j \in \mathcal{C}}\sqrt{1-\frac{(\omega_j-\Omega)^2}{K^2r^2}}\label{resultant_eqn_cc_arcsin_1}\\
r\sin\lambda&=\frac{1}{N}\left(\sum_{j \in \mathcal{C}}\frac{\omega_j-\Omega}{Kr}+\sum_{j \notin \mathcal{C}}k_j\right). \label{resultant_eqn_cc_arcsin_2}
\end{align}
In the thermodynamic limit $N\rightarrow \infty$ this recovers the mean-field self consistency equations (\ref{self_consistency_real})--(\ref{self_consistency_imag}), as shown in Appendix \ref{sec:reduced_equation}. This correspondence with the mean-field theory is only achieved for the arcsine ansatz and when the rogue oscillators are taken into account.\\

In Sec.~\ref{sec:results} we further show that the collective coordinate equations using the arcsine ansatz capture the collective dynamics of the full Kuramoto-Sakaguchi model with finitely many oscillators more accurately than when using the linear ansatz. However we remark that the linear ansatz (\ref{ansatz_linear}), defined here for an all-to-all coupling network, can be extended to arbitrary network topologies, while the (more accurate) arcsine ansatz (\ref{ansatz_arcsine}) is restricted to a globally connected all-to-all network.\\

At this stage, we have tacitly assumed that the cluster is known and we can separate the oscillators into those that synchronize $i\in\mathcal{C}$ and the non-entrained rogues $i\notin \mathcal{C}$. To obtain the cluster via the collective coordinate method, we assume that if nodes can synchronize, they will do so. We therefore seek the maximal set of synchronized oscillators $\mathcal{C}$ such that the system of reduced algebro-differential equations (\ref{reduced_equation_a})--(\ref{reduced_equation_b}) has a stable stationary solution. If stationary solutions cannot be found, we exclude nodes $i$ with maximal value of $|\Omega-\omega_i|$, check again for existence of stationary solutions, and, if needed, repeat this procedure until we arrive at a set $\mathcal{C}$ for which a stationary solution can be found. This criterion, however, is not sufficient to find the best approximation for the synchronized cluster, as there may be a stationary solution of (\ref{reduced_equation_a})--(\ref{reduced_equation_b}) which is linearly stable within the ansatz manifold $\Theta_i(r,\Omega)$, but is unstable in the full Kuramoto-Sakaguchi model (\ref{KS-theta}), that is, the dynamics transverse to the ansatz manifold near the stationary solution is unstable. To account for this, we study the stability of the approximated phases $\Theta_i$ by substituting $\theta_i(t)=\Theta_i+\eta_i(t)$, where $\eta_i(t)$ represents small perturbations, into the full Kuramoto-Sakaguchi model (\ref{KS-theta}). Expanding up to linear order of $\eta$ and assuming $\Theta_i$ satisfies the equation at lowest order, we arrive at a linear system of the form
\begin{equation} 
\dot{\eta}_i=\sum_{j \in \mathcal{C}}L_{ij}\eta_j,
\end{equation}
where
\begin{widetext}
\begin{equation}  
L_{ij}= \begin{cases}\cos(\Theta_j-\Theta_i-\lambda), \hspace{5pt} &j \neq i\\
-\sum\limits_{l \in \mathcal{C},l \neq i}\cos(\Theta_l-\Theta_i-\lambda) -\sin(\Theta_i+2\lambda)\sum\limits_{l \notin \mathcal{C}}k_l, \hspace{5pt} &j=i
\end{cases}	
.				
\end{equation}
\end{widetext}
%The matrix $L$ always has an eigenvector $(1,1,\dots,1)$ with eigenvalue $\lambda_1=0$, corresponding to the system's invariance to a constant phase shift. If all other eigenvectors have eigenvalues with negative real parts, 
%If all eigenvectors of $L$ have eigenvalues with non-positive real parts, 
The matrix $L$ always has an eigenvector $(1,1,\dots,1)$ with eigenvalue $\lambda_1=0$, corresponding to the system's invariance to a constant phase shift, if all oscillators partake in synchronisation. The presence of rogue oscillators leads to perturbations of this eigenvector and its associated eigenvalue. If all other eigenvectors have eigenvalues with negative real parts, we consider the phases $\Theta_i$ and the stationary solution $r$ and $\Omega$ to be stable in the full Kuramoto-Sakaguchi model. Note that this assumes that our collective coordinate ansatz and the stationary solutions $r$ and $\Omega$ are indeed a good approximation of the actual phases $\theta_i$. If $\Theta_i$ is not stable according to this definition, we again exclude the node with intrinsic frequency having maximal $|\Omega-\omega_i|$. This procedure is repeated until a stable stationary solution of the reduced equations (\ref{reduced_equation_a})--(\ref{reduced_equation_b}) is found. At each step we check that there are no stable solutions possible for nearby clusters, displaced by up to $5$ nodes.

Given an ansatz function $\Theta$, i.e., the linear ansatz (\ref{ansatz_linear}) or the arcsine ansatz (\ref{ansatz_arcsine}), we can now determine the order parameter $\bar r$, the cluster mean frequency $\Omega$, the size of the synchronized cluster and other properties of the collective behavior. When the effect of the rogue oscillators on the synchronized cluster is taken into account, the order parameter $r$ obtained as the stationary solution of the collective coordinate evolution equation (\ref{reduced_equation_c}) should satisfy
\begin{equation} {\label{rbar_approximation}}
r = \left|\frac{1}{N}\left(\sum_{j \in \mathcal{C}}e^{i \Theta_j}+\sum_{j \notin \mathcal{C}}\int_0^{2 \pi}e^{i \theta_j} P_j(\theta_j)d \theta_j\right)\right|,
\end{equation}
where $P_j(\theta_j)$ is the probability density function of the rogue oscillators (\ref{ansatz_rogue_distribution}). This equality, however, is only ensured for the arcsine ansatz and only if the rogue oscillators are included in the collective coordinate approach. For the arcsine ansatz (\ref{ansatz_arcsine}) and the distributional ansatz for the rogue oscillators (\ref{ansatz_rogue_distribution}), splitting the real and imaginary parts of the sum within the absolute value readily yields (\ref{resultant_eqn_cc_arcsin_1})--(\ref{resultant_eqn_cc_arcsin_2}), respectively. For the linear ansatz function the equality is only approximately satisfied up to $\mathcal{O}(1/K^2)$.\\

In the following section we demonstrate how each of the two ansatz functions (\ref{ansatz_linear}) and (\ref{ansatz_arcsine}) performs in reproducing the collective behavior of the full Kuramoto-Sakaguchi model, and, in particular, we show how the inclusion of the rogue oscillators is necessary to accurately capture the dynamics of the partially synchronized state near the onset of synchronization.

%%%%%%%%%%%%%%%%%%%%%%%%%%%%%%%%%%%%%%%%%%%

\section{Numerical Results} 
\label{sec:results}

In this section we illustrate the efficacy of the collective coordinate approach with the linear ansatz (\ref{ansatz_linear}) and the arcsine ansatz (\ref{ansatz_arcsine}) to approximate the collective behavior of the full Kuramoto-Sakaguchi model (\ref{KS-theta}). We will cast particular emphasis on the inclusion or neglect of the effect of the rogue oscillators. Specifically, we present estimates for the order parameter $\bar{r}$ and the mean cluster frequency $\Omega$, as well as identification of the synchronized cluster $\mathcal{C}$ and the critical coupling strength for partial synchronization $K_c$ and for global synchronization $K_g$. In the collective coordinate framework, the approximation for $\Omega$ is obtained directly from the stable stationary solution of the reduced equations (\ref{reduced_equation_a})--(\ref{reduced_equation_b}), and the approximation for $\bar r$ is obtained by substituting the stationary solution into (\ref{rbar_approximation}). We note that for the arcsine ansatz with rogues included, $\bar r$ obtained from  (\ref{rbar_approximation}) coincides with the stationary solution of the order parameter of (\ref{reduced_equation_a})--(\ref{reduced_equation_b}), i.e., it is self consistent, as discussed in the previous section. For the linear ansatz we found that using $r$ obtained from (\ref{rbar_approximation}) yields a more accurate approximation of the order parameter than using the stationary solution directly. Note that in the cases where the rogue oscillators are ignored when finding the solution to (\ref{reduced_equation_a})--(\ref{reduced_equation_b}), we still include the influence of rogues in (\ref{rbar_approximation}). We find there is only a small difference between including or ignoring the rogue oscillators in (\ref{rbar_approximation}) in these cases.

Before we embark on our numerical study, we describe briefly the parameters used for numerical simulation of the full Kuramoto-Sakaguchi model (\ref{KS-1}). We use random initial conditions with a fourth-order Runge-Kutta method for at least $2000$ time units to ensure convergence of the order parameter. The first half is discarded to discard transient behavior.

Real stationary solutions $r$ and $\Omega$ of the collective coordinate approximation (\ref{reduced_equation_a})--(\ref{reduced_equation_b}) are found numerically using MATLAB's \texttt{fsolve} function \cite{MATLAB}. The algebro-differential equations for the collective coordinates (\ref{reduced_equation_a})--(\ref{reduced_equation_b}) typically have a pair of solutions, one stable and one unstable, corresponding to a saddle-node bifurcation, and care needs to be taken to select the correct stable solution. This is achieved by maintaining continuity when $K$ is varied. Furthermore, (\ref{reduced_equation_a})--(\ref{reduced_equation_b}) contain square roots and arcsine functions and we need to check that their arguments are within the respective domain. 

A brief comment on the computational cost of the collective coordinate method compared to direct numerical simulation of the Kuramoto-Sakaguchi model. To estimate the order parameter using the collective coordinate method there are several steps involved. The root finding method to obtain the stationary solutions of the collective coordinates involves fast standard root finding routines such as \texttt{fsolve} in MATLAB \cite{MATLAB}. When cycling through different values of the coupling strength the stationary solutions of a previous value of the coupling strength can be used as initial condition for the adjacent coupling strength to facilitate the root finding. In the case that no stationary solution can be found, we check whether a stationary solution exists for synchronized clusters shifted up to 5 frequencies in each direction. The stability of stationary solutions that are found is then tested by employing standard eigenvalue routines for the Jacobian. The computational cost has to be set against the computational complexity associated with the temporal evolution of the full Kuramoto-Sakaguchi model. Since the Kuramoto-Sakaguchi model does not evolve into stationary states due to the effect of the rogue oscillators, the length of the simulation may be very large to ensure convergence of the averaged order parameter. This is especially the case close to the bifurcation from incoherence to partial synchronization when the number of rogue oscillators is large this can be prohibitive.\\

%%%%%%%%%%%%%%%%%%%%%%%%%%%%%%%%%%%%%%%%%%%

\subsection{Order parameter $\bar{r}$}

\begin{figure}[tbh]
	\centering
	\includegraphics[width=0.5\textwidth]{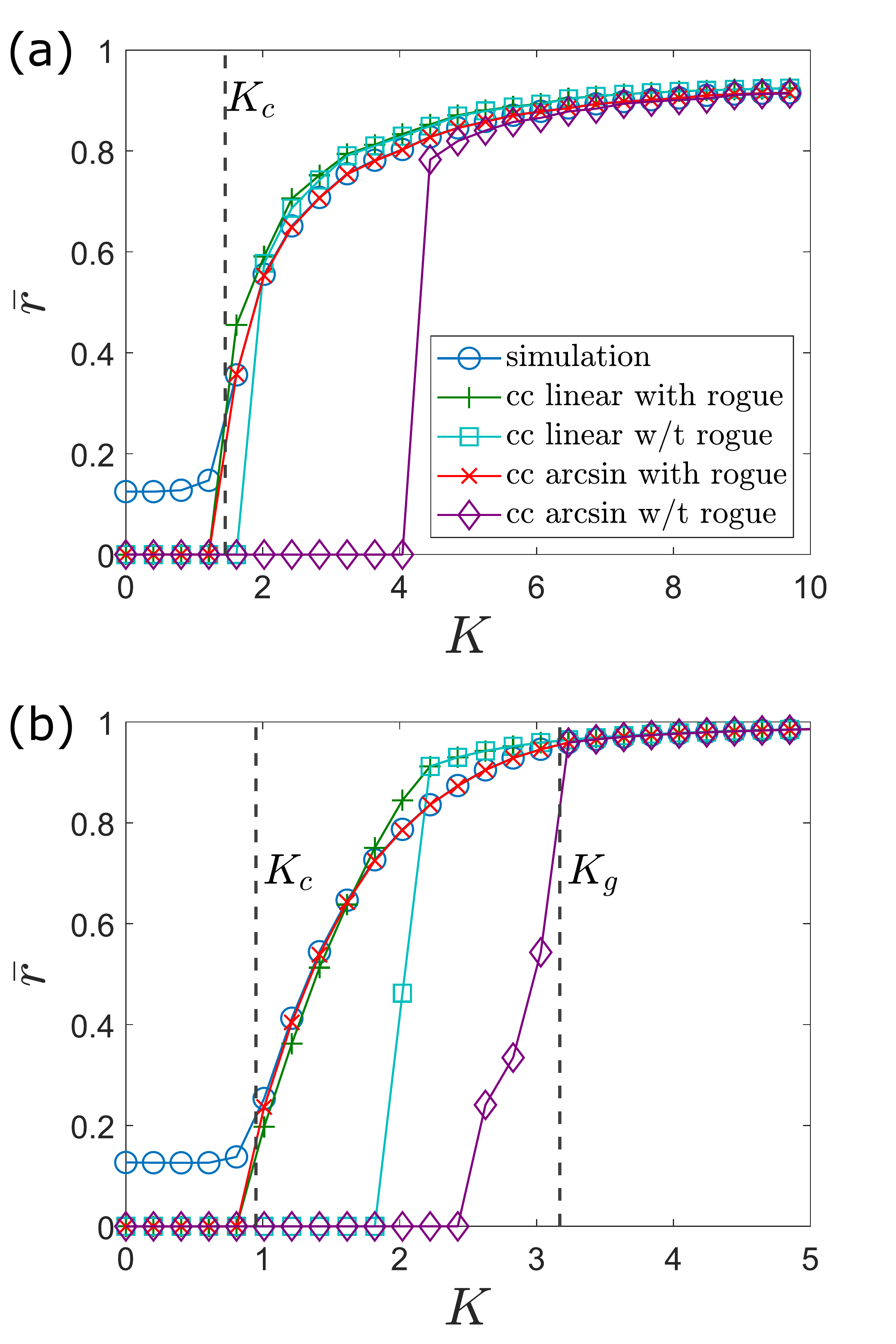}
	\caption{Order parameter $\bar{r}$ for the Kuramoto-Sakaguchi model (\ref{KS-1}) with $\lambda=\pi/4$ and $N=50$ oscillators. Shown are estimates obtained using the collective coordinate approach (labelled cc) with a linear ansatz (\ref{ansatz_linear}) and an arcsine ansatz (\ref{ansatz_arcsine}), both with and without including the effect of rogue oscillators, as well as obtained from a numerical simulation of the full Kuramoto-Sakaguchi model (\ref{KS-1}). (a): Lorentzian distribution (\ref{lorentzian_distribution}). (b): uniform distribution (\ref{uniform_distribution})}
	\label{fig:comparison_r} 	
\end{figure}

\begin{figure*}[tbh]
	\centering
	\includegraphics[width=\textwidth]{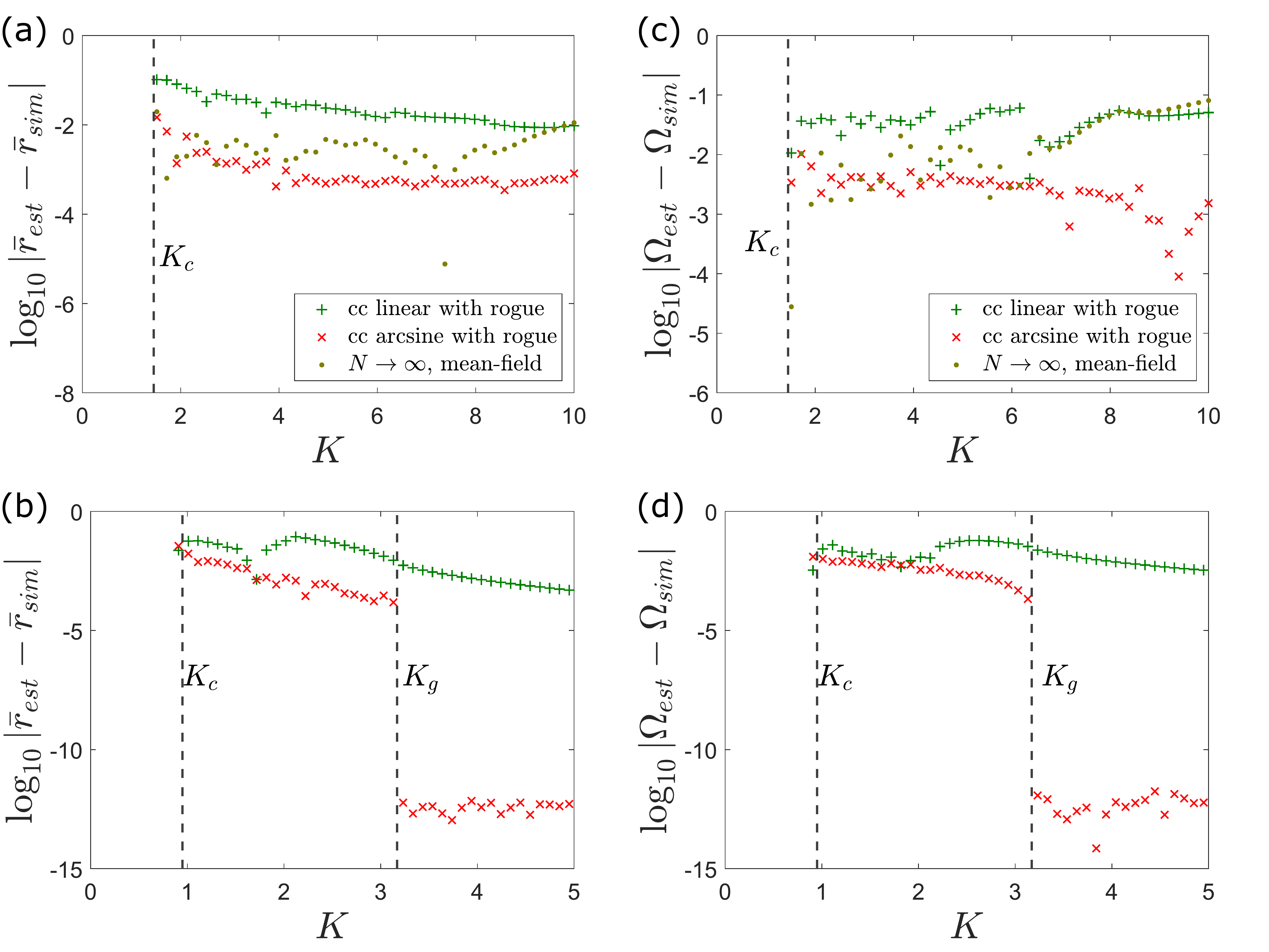}
	\caption{Error in estimates of the order parameter $\bar{r}$ (a,b) and of the cluster mean frequency $\Omega$ (c,d) for the various collective coordinate approaches (labelled cc)  shown in Fig.~\ref{fig:comparison_r}. Top row: Lorentzian distribution (\ref{lorentzian_distribution}); shown is also the mean-field result (\ref{self_consistency_real})--(\ref{self_consistency_imag}) for comparison. Bottom row: uniform distribution (\ref{uniform_distribution}).}
	\label{fig:comparison} 	
\end{figure*}

Fig.~\ref{fig:comparison} shows the order parameter $\bar{r}$ for a phase-frustration parameter $\lambda=\pi/4$. We show results obtained from the collective coordinate approach, where $\bar r$ is defined via (\ref{rbar_approximation}), using both the linear ansatz (\ref{ansatz_linear}) and the arcsine ansatz (\ref{ansatz_arcsine}), and both with and without the inclusion of the rogue oscillators. This is compared with the order parameter obtained from a numerical simulation of the full Kuramoto-Sakaguchi model (\ref{KS-1}). We show results for a Lorentzian intrinsic frequency distribution (\ref{lorentzian_distribution}) (Fig.~\ref{fig:comparison_r}(a)) and for a uniform intrinsic frequency distribution (\ref{uniform_distribution}) (Fig.~\ref{fig:comparison_r}(b)).

It is seen that including the rogue oscillators is crucial in obtaining a correspondence between the results of the full Kuramoto-Sakaguchi model (\ref{KS-1}) and of our model reduction, for both frequency distributions and for both types of ansatz. The effect the rogue of oscillators is particularly striking for the arcsine ansatz; including the rogues leads to a very small error in the estimate of the order parameter for all values of the coupling strength $K$, whereas without the rogue oscillators no stationary solutions can be found for $K\lesssim 4$ for the Lorentzian distribution and for $K\lesssim2.5$ for the uniform distribution, completely missing the critical coupling strengths for the onset of partial synchronization at $K_c\approx1.45$ and $K_c\approx 0.95$, for the respective intrinsic frequency distributions. Similarly, the inclusion of rogue oscillators markedly improves the estimate of the order parameter for the linear ansatz.

To quantify the accuracy of the collective coordinate approaches, we compute the error in their estimates of $\bar{r}$, when the rogue oscillators are included, compared to $\bar{r}$ obtained from a simulation of the full Kuramoto-Sakaguchi model (\ref{KS-1}). We also compute the order parameter $\bar{r}$ using the Ott-Antonsen ansatz in the thermodynamic limit $N\rightarrow\infty$ for Lorentzian frequency distributions (explicit expressions 
%We also compute the order parameter $\bar{r}$ obtained from the mean-field analysis in the thermodynamic limit $N\rightarrow\infty$ as solution to (\ref{self_consistency_real})--(\ref{self_consistency_imag}) for Lorentzian frequency distributions (explicit expressions % for the Lorentzian %and the uniform 
%intrinsic frequency distributions 
are provided in Appendix \ref{sec:thermodynamics_1}). The errors are plotted in Fig.~\ref{fig:comparison}(a,b) for the Lorentzian and the uniform frequency distribution, respectively. The error is lowest for the arcsine ansatz which is designed for finite networks. For the uniform frequency distribution, the error of the arcsine ansatz is of the order of the numerical round-off error when the system is in global synchronization with $K>K_g\approx 3.2$; this is because for global synchronization the arsine ansatz (\ref{ansatz_arcsine}) is exact. For the Lorentzian frequency distribution $K_g \approx 53.7$ and is not in the range shown in Fig.~\ref{fig:comparison_r}. The mean-field solution is generally more accurate than the collective coordinate solution using the linear ansatz. The sharp drops of the error 
%of the collective coordinate estimate using the linear ansatz and 
of the mean-field limit for the Lorentzian frequency distribution stem from the estimate of $\bar r(K)$ crossing the curve of the order parameter of the full Kuramoto-Sakaguchi model 
%near $K\approx6.8$ for the linear ansatz and near $K\approx3.8$
near $K\approx7.5$ for the mean-field limit. % (cf. Fig.~\ref{fig:comparison}(a)).

%%%%%%%%%%%%%%%%%%%%%%%%%%%%%%%%%%%%%%%%%%%

\subsection{Cluster mean frequency $\Omega$}

Fig.~\ref{fig:comparison}(c,d) shows the error in estimating the cluster mean frequency $\Omega$ for a phase-frustration parameter $\lambda=\pi/4$ (cf. Fig.~\ref{fig:Omega_vs_K} for $\Omega(K)$ for the full Kuramoto-Sakaguchi model). We show the error for the collective coordinate approach, for the linear ansatz (\ref{ansatz_linear}) and the arcsine ansatz (\ref{ansatz_arcsine}), both with the inclusion of the rogue oscillators. The cluster mean frequency $\Omega$ for the collective coordinates is again obtained as the stationary solution $\Omega$ of the reduced equations (\ref{reduced_equation_a})--(\ref{reduced_equation_b}). We further show the error of the Ott-Antonsen ansatz 
%of the mean-field analysis as solution to (\ref{self_consistency_real})--(\ref{self_consistency_imag}) 
for the Lorentzian frequency distribution. Again, the collective coordinate approach using the arcsine ansatz achieves the smallest error, and since the ansatz is exact for globally synchronized oscillators the error is negligible for $K>K_g$. 

Not surprisingly, the behavior of the error for $\Omega$ echos the behavior of the error for the order parameter. As for the order parameter $\bar{r}$, including the effect of the rogue oscillators is crucial to accurately estimate the cluster mean frequency of the actual Kuramoto-Sakaguchi model (\ref{KS-1}) for the whole range of coupling strengths. When the effect of the rogue oscillators is included, the cluster mean frequency $\Omega(K)$ is very well approximated by the collective coordinate approach, for both the linear and the arcsine ansatz. In particular, the non-monotonic dependence of $\Omega$ on $K$ for the uniform distribution near the onset of partial synchronization as observed in Fig.~\ref{fig:Omega_vs_K}(b) is well captured (not shown). Recall that for the range of coupling strengths $K$ used here, the onset of global synchronization is only depicted for the case of uniformly distributed frequencies. Again, the collective coordinate approach using the linear ansatz exhibits the largest errors, except for dips when the curves $\Omega(K)$ obtained from the collective coordinates crosses the curve of the full Kuramoto-Sakaguchi model.

%%%%%%%%%%%%%%%%%%%%%%%%%%%%%%%%%%%%%%%%%%%

\subsection{Synchronized cluster $\mathcal{C}$}

\begin{figure} 
	\centering
	\includegraphics[width=1\linewidth]{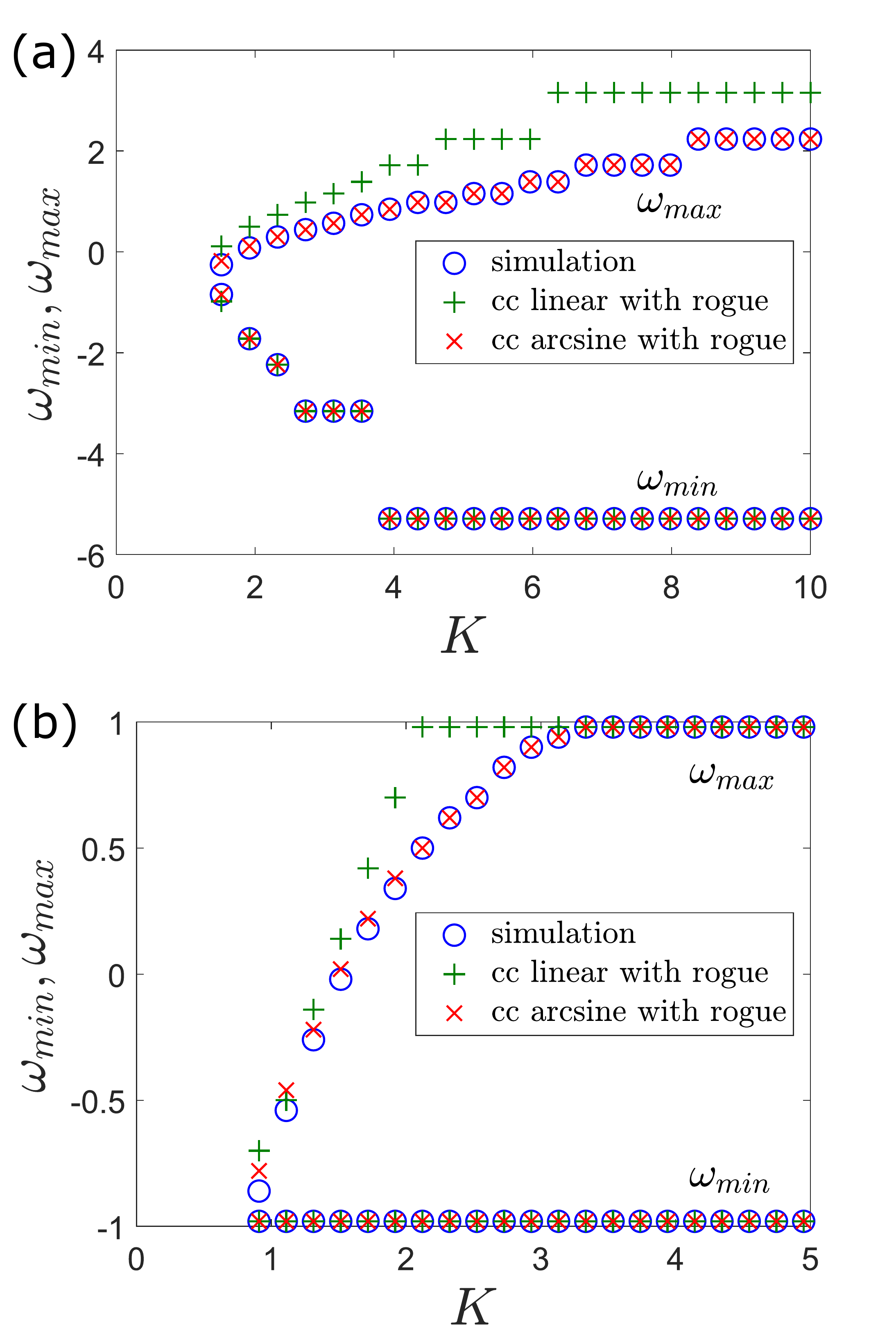}
	\caption{Minimal and maximal intrinsic frequencies ($\omega_{\text{min}},\omega_{\text{max}}$) of the synchronized cluster $\mathcal{C}$ for a phase-frustration $\lambda=\pi/4$. We show results obtained from the collective coordinate approach (labelled cc) using the linear ansatz (\ref{ansatz_linear}) and the arcsine ansatz (\ref{ansatz_arcsine}), both including the effect of rogue oscillators, and obtained from a numerical simulation of the full Kuramoto-Sakaguchi model (\ref{KS-1}) with $N=50$ oscillators. (a) Lorentzian frequency distribution (\ref{lorentzian_distribution}). (b) Uniform frequency distribution (\ref{uniform_distribution}).}
	\label{fig:N12_comparison}
\end{figure}

Fig.~\ref{fig:N12_comparison} shows the minimal and maximal intrinsic frequencies $\omega_{\text{min}}$ and $\omega_{\text{max}}$, respectively, such that all oscillators with intrinsic frequencies $\omega_{\min}\leq \omega_i \leq \omega_{\max}$ partake in synchronized dynamics. The rogue oscillators, by definiton, are those with intrinsic frequencies outside this range (ie with intrinsic frequencies $\omega_i<\omega_{\text{min}}$ or $\omega_i>\omega_{\text{max}}$). We show results of the synchronized cluster obtained from numerical simulation of the full Kuramoto-Sakaguchi model (\ref{KS-1}), as well as results from the collective coordinate approach using the linear ansatz (\ref{ansatz_linear}) and the arcsine ansatz (\ref{ansatz_arcsine}), both with the inclusion of the rogue oscillators. We recall that, for the collective coordinate approach, the synchronized cluster $\mathcal{C}$ is defined as the largest set of oscillators for which the reduced equations (\ref{reduced_equation_a})--(\ref{reduced_equation_b}) have a stationary solution that is stable in the full Kuramoto-Sakaguchi model. It is seen that the arcsine ansatz captures the synchronized cluster very well, with small discrepancies occurring only close to the onset of partial synchronization at $K_c$. The linear ansatz tends to overpredict the size of the synchronized cluster. 
%Our method is implemented to search for the maximal cluster such that a stationary solution of the reduced equation (\ref{reduced_equation_1}) exists and is stable in the full Kuramoto-Sakaguchi model (\ref{KS-theta}). For the linear ansatz, such stable stationary solution in certain cases exists for a multiple of different clusters and our implementation picks up the largest one, which is sometimes different to the synchronized cluster observed in numerical simulation. For the arcsine ansatz, on the other hand, stable stationary solution of (\ref{reduced_equation_1}) can be found at fewer clusters and the largest cluster tend to agree with the synchronized cluster observed in numerical simulation, although near the onset of partial synchronization the arcsine ansatz sometimes also over-predicts.

%%%%%%%%%%%%%%%%%%%%%%%%%%%%%%%%%%%%%%%%%%%

\subsection{Critical coupling strengths $K_c$ and $K_g$}

We estimate the critical coupling strength $K_c=K_c(\lambda)$ corresponding to the onset of partial synchronization as the smallest value of $K$ such that the order parameter $\bar{r}$ exceeds a threshold value $\bar r>0.2$, where $\bar r(K)$ is sampled in increments $\Delta K=0.01$. We obtain $K_c(\lambda)$ in this way for the full Kuramoto-Sakaguchi model (\ref{KS-1}) and for the collective coordinate approach with the arcsine ansatz and rogues included (\ref{reduced_equation_a})--(\ref{reduced_equation_b}). %, and for the mean-field analysis (\ref{self_consistency_real})--(\ref{self_consistency_imag}) which is valid in the thermodynamic limit. 
The critical coupling strength $K_g$ corresponding to the onset of global synchronization is obtained such that all of the oscillators synchronize. For the collective coordinate approach, $K_g$ is defined as the lowest value of $K$ such that a stationary solution of the reduced equations (\ref{reduced_equation_a})--(\ref{reduced_equation_b}) exists for $\mathcal{C}$ consisting of all $N$ oscillators, and is stable in the full Kuramoto-Sakaguchi model (\ref{KS-theta}).

\begin{figure} 
	\centering
	\includegraphics[width=1\linewidth]{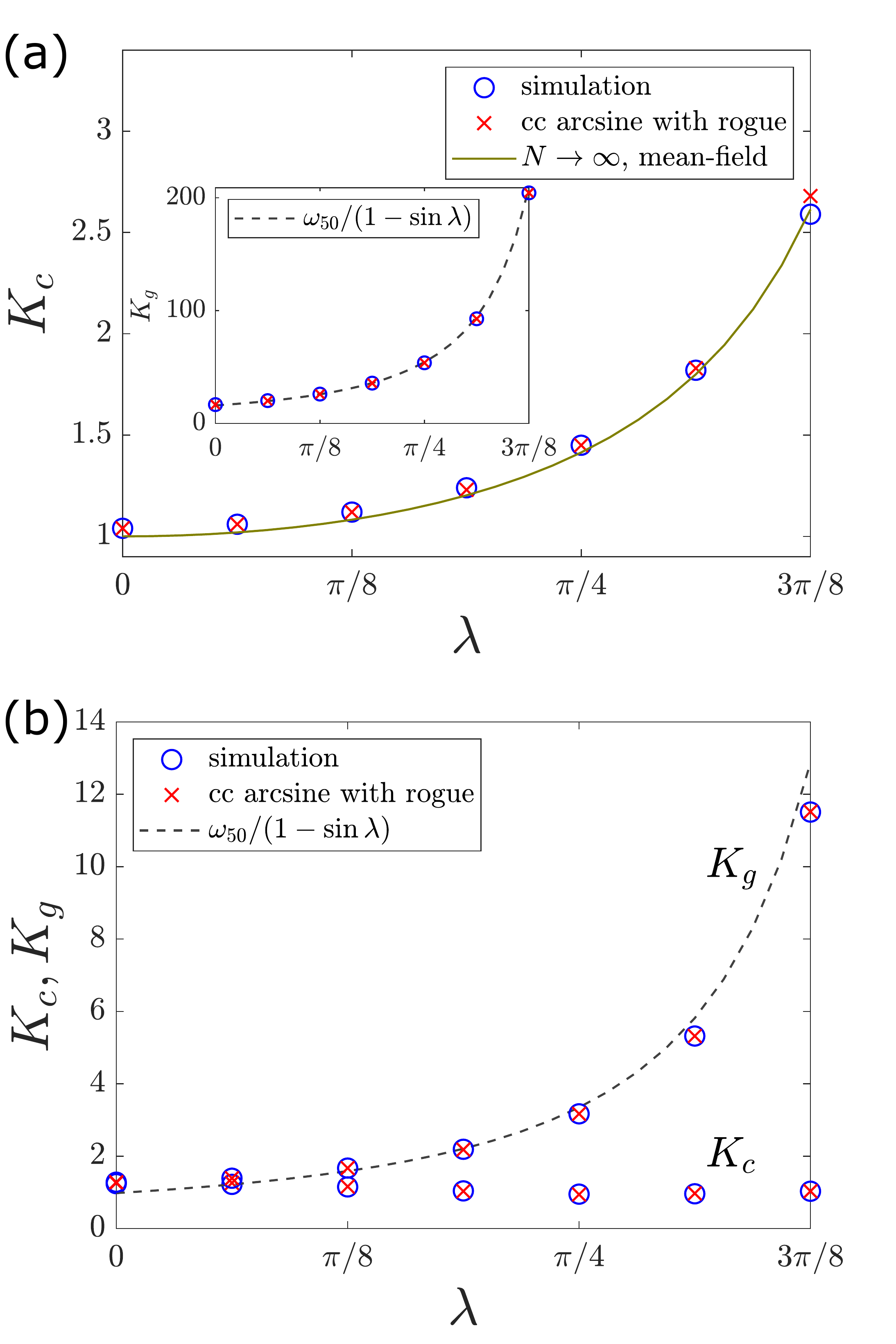}
%	\caption{Critical coupling strengths $K_c$ and $K_g$ obtained from a numerical simulation of the full Kuramoto-Sakaguchi model (\ref{KS-1}) with $N=50$ oscillators, and obtained from the collective coordinate approach using the the arcsine ansatz (\ref{ansatz_arcsine}) including the effect of rogue oscillators, as well as obtained from the mean-field result (\ref{self_consistency_real})--(\ref{self_consistency_imag}). (a) Lorentzian frequency distribution (\ref{lorentzian_distribution}). (b) Uniform frequency distribution  (\ref{uniform_distribution}).}
	\caption{Critical coupling strengths $K_c$ and $K_g$ obtained from a numerical simulation of the full Kuramoto-Sakaguchi model (\ref{KS-1}) with $N=50$ oscillators, and obtained from the collective coordinate approach (labelled cc)  using the the arcsine ansatz (\ref{ansatz_arcsine}) including the effect of rogue oscillators. For the onset of global synchronization we also show the estimate $K_g=\omega_{50}/(1-\sin \lambda)$ borrowed from a mean-field analysis. For the onset of partial synchronization we also show results from the Ott-Antonsen ansatz for the Lorentzian frequency distribution.  (a) Lorentzian frequency distribution (\ref{lorentzian_distribution}). (b) Uniform frequency distribution  (\ref{uniform_distribution}).}
	\label{fig:kc_comparison}
\end{figure}

In Fig.~\ref{fig:kc_comparison} we compare the critical coupling strength $K_c$ and $K_g$ as a function of the phase-frustration parameter $\lambda$ estimated from numerical simulations of the full Kuramoto-Sakaguchi model (\ref{KS-1}) with $N=50$ oscillators, and from the collective coordinate approach (using the arcsine-ansatz including the rogue oscillators) for the Lorentzian and the uniform frequency distribution. For both frequency distributions the collective coordinate approach captures the onset of partial and of global synchronization remarkably well. For the Lorentzian frequency distribution we also show results of the mean-field analysis for partial synchronization which captures the onset very well even for the finite network with only $N=50$ oscillators. The onset of global synchronization can be approximated using the mean-field analysis relation $\Omega=-Kr^2\sin \lambda$ (cf. (\ref{e.OmKr})). Approximating $r\approx1$ and that the last oscillator to be entrained at the onset of global synchronization is $\omega_N$, we approximate $K_g=\omega_{50}/(1-\sin \lambda)$. This approximation also captures the onset of global synchronization very well.  

%, as well as from the mean-field analysis in thermodynamic limit. For the Lorentzian frequency distribution we only depict the critical coupling strength $K_c$ for the onset of partial synchronization (Fig.~\ref{fig:kc_comparison}(a)). The onset of global synchronization $K_g$ occurs at much larger values of the coupling strength is not shown. 

%For the uniform frequency distribution, both $K_c$ and $K_g$ are depicted (Fig.~\ref{fig:kc_comparison}(b)). For both frequency distributions, the collective coordinate approach with the arcsine ansatz including the effect of the rogue oscillators and the mean-field analysis reproduce the onsets of partial and global synchronization remarkably well, with the collective coordinate method more accurately capturing the critical values. The estimate of the onset of global synchronization $K_g$ from the mean-field analysis deviates from the truth for $\lambda>\pi/4$ and mean-field analysis predicts a delayed onset of global synchronization for the uniform frequency distribution.
%For the uniform frequency distribution we also see that for $\lambda=0$ $K_c=K_g$, i.e., partial synchronization does not occur for $\lambda=0$. This is shown analytically for the collective coordinate approach in Appendix~\ref{sec:thermodynamic_uniform}.

%$K_g\approx53.7$

%%%%%%%%%%%%%%%%%%%%%%%%%%%%%%%%%%%%%%%%%%%

\section{Summary and outlook} \label{sec:conclusion}

We have derived reduced dynamics of the Kuramoto-Sakaguchi model through the collective coordinate framework. We have extended the collective coordinate approach by including the effect of non-entrained rogue oscillators, and have shown that including these rogue oscillators is essential to accurately describe the collective dynamics of the Kuramoto-Sakaguchi model. We have compared two ansatz functions, a linear ansatz function obtained as a linearization around $1/K$, and a nonlinear arcsine-ansatz, motivated by mean-field analysis. We find that the arcsine ansatz with the effect of the rogues included provides a remarkably good approximation of the collective dynamics of finite networks, quantified by the order parameter, the cluster mean frequency, the identification of the cluster as well as the critical coupling strengths for partial and for global synchronization, and we find that the arcsine ansatz is far superior to the linear ansatz. However, the arcsine ansatz is limited to networks with an all-to-all coupling topology, whereas the linear ansatz can be applied to any complex network \cite{HancockGottwald_2018}. We have also shown that for finite networks the arcsine ansatz is far superior to classical self-consistency equations, which assume $N\to \infty$. Moreover, in the thermodynamic limit of infinitely many oscillators, we have shown that the arcsine ansatz recovers well-known results obtained by mean-field analysis.\\

Taken the issue of computational cost aside achieved by the reduction from $N$ oscillators to $2$ collective coordinates, the advantage of the model reduction presented here is a reduced dynamical description allows a more detailed analysis of the macroscopic dynamics; in particular, to establish and quantitatively capture dominant effects such as the influence of the mean-field of the non-entrained rogue oscillators on the synchronized macroscopic behavior.\\   

Here we have considered the Kuramoto-Sakaguchi model with an all-to-all coupling and a phase-frustration parameter that is common to all oscillators. Generalization of the model, such as a system consisting of two populations of oscillators with different inter- and intra-population coupling strengths and phase-frustrations, have been shown to display intriguing dynamics such as chimera states and chaos \cite{AbramsEtAl_2008, Laing_Chaos2009, BickEtAl18}. The success of the collective coordinate approach in the one-population Kuramoto-Sakaguchi model suggests that it will also be able to capture the complex collective dynamics of those more general models. Furthermore, the methods developed in \cite{HancockGottwald_2018,SmithGottwald_2019} allow to study local frequency clusters, caused by finite size sampling effects in the natural frequencies, and their mutual interaction in phase-frustrated systems. 

\begin{acknowledgments}
We wish to acknowledge support from the Australian Research Council, Grant No. DP180101991.
\end{acknowledgments}

%%%%%%%%%%%%%%%%%%%%%%%%%%%%%%%%%%%%%%%%%%%

\appendix 
\section{Reduced Equations of the Collective Coordinate Approach for the Linear and the Arcsine Ansatz} 
\label{sec:reduced_equation}

We recall the condition for the minimization of the error (\ref{eq:errormin}):
\begin{equation} {\label{cc_orthogonal_1}}
\sum_{i\in\mathcal{C}}\mathcal{E}_i\frac{\partial \Theta_i}{\partial r}=0\hspace{5pt} \text{and} \hspace{5pt}\sum_{i \in \mathcal{C}}\mathcal{E}_i\frac{\partial \Theta_i}{\partial \Omega}=0.
\end{equation}
\noindent
{\bf{Linear ansatz:}} For the linear ansatz (\ref{ansatz_linear}) we evaluate $\frac{\partial \Theta_i}{\partial r}=-\frac{\omega_i-\Omega}{Kr^2}$, $\frac{\partial \Theta_i}{\partial \Omega}=-\frac{1}{Kr}$, and (\ref{cc_orthogonal_1}) leads to the following system of two equations
\begin{equation} 
\dot{r}\bm{H}_{\text{lin}}=\bm{F}_{\text{lin}}(r,\Omega),
\label{eq:ApA2728}
\end{equation}
where
\begin{align*} 
\bm{H}_{\text{lin}}&=\begin{pmatrix} \sum\limits_{i \in \mathcal{C}} \frac{(\omega_i-\Omega)^2}{K^2r^4} \\ \sum\limits_{i \in \mathcal{C}} \frac{\omega_i-\Omega}{K^2r^3}  \end{pmatrix},\\
\bm{F}_{\text{lin}}&=-\begin{pmatrix}\sum\limits_{i \in \mathcal{C}} \frac{(\omega_i-\Omega)^2}{Kr^2}+\frac{K}{N}\sum\limits_{i \in \mathcal{C}}\frac{\omega_i-\Omega}{Kr^2}h_i\\ \sum\limits_{i \in \mathcal{C}} \frac{(\omega_i-\Omega)}{Kr}+\frac{K}{N}\sum\limits_{i \in \mathcal{C}}\frac{1}{Kr}h_i\end{pmatrix},
\end{align*}
with
\begin{align*} 
h_i=\sum_{j \in C}\sin(\Theta_j-\Theta_i-\lambda)+\cos(\Theta_i+2 \lambda) \sum_{j \notin C}k_j,
\end{align*}
where $k_j$ are as in (\ref{k_j}). Simplifying (\ref{eq:ApA2728}) leads to (\ref{reduced_equation_a})--(\ref{reduced_equation_b}).\\

\noindent
{\bf{Arcsine ansatz:}} Defining
\begin{align*} 
s_i=\frac{\omega_i-\Omega}{Kr} \hspace{5pt} \text{and} \hspace{5pt} c_i=\sqrt{1-\frac{(\omega_i-\Omega)^2}{K^2r^2}},
\end{align*}
we evaluate for the arcsine ansatz (\ref{ansatz_arcsine}) 
$\frac{\partial \Theta_i}{\partial r}=-\frac{s_i}{r c_i}$, $\frac{\partial \Theta_i}{\partial \Omega}=-\frac{1}{Kr c_i}$, and (\ref{cc_orthogonal_1}) leads to the following system of two equations
\begin{equation} 
\dot{r}\bm{H}_{\text{asin}}=\bm{F}_{\text{asin}}(r,\Omega),
\label{eq.AppA_2728asin}
\end{equation}
where
\begin{align*} 
\bm{H}_{\text{asin}}&=\begin{pmatrix} \frac{1}{r^2} \sum_{i \in \mathcal{C}}\frac{s_i^2}{c_i^2}\\ \frac{1}{Kr^2}\sum_{i \in \mathcal{C}}\frac{s_i}{c_i^2} \end{pmatrix}\\
\bm{F}_{\text{asin}}&=-G\begin{pmatrix}F_1\\F_2\end{pmatrix}\\
G&=\begin{pmatrix} KC &KA \\ E &N_c\end{pmatrix}\\
F_1&=1-\frac{1}{Nr}(A\sin\lambda+B\cos\lambda+D\sin\lambda)\\
F_2&=\frac{1}{Nr}(A\cos\lambda-B\sin\lambda+D\cos\lambda),
\end{align*}
with
\begin{align*} 
A&=\sum_{i \in \mathcal{C}}s_i,\hspace{5pt} B=\sum_{i \in \mathcal{C}}c_i,\hspace{5pt} C=\sum_{i \in \mathcal{C}}\frac{s_i^2}{c_i},\\
D&=\sum_{i \notin \mathcal{C}}k_i,\hspace{5pt} E=\sum_{i \in \mathcal{C}}\frac{s_i}{c_i}.
\end{align*}
Simplifying (\ref{eq.AppA_2728asin}) leads to (\ref{reduced_equation_a})--(\ref{reduced_equation_b}).

For the arcsine ansatz, to look for stationary solutions of the reduced equation, we set $\dot{r}=0$, which results in
\begin{equation*} 
\bm{0}=\bm{F}_{\text{asin}}(r,\Omega).
\end{equation*}
This equation is satisfied if $F_1=0$, $F_2=0$, which is equivalent to
\begin{align*} 
r\cos\lambda&=\frac{1}{N}\sum_{j \in \mathcal{C}}c_j,\\
r\sin\lambda&=\frac{1}{N}\left(\sum_{j\in\mathcal{C}}s_j+\sum_{j\notin\mathcal{C}}k_j\right).
\end{align*}

In the thermodynamic limit $N\rightarrow\infty$, the above equations recover the self-consistency result derived from mean-field analysis  (\ref{self_consistency_real})--(\ref{self_consistency_imag}).

%%%%%%%%%%%%%%%%%%%%%%%%%%%%%%%%%%%%%%%%%%%

\section{The Kuramoto-Sakaguchi Model in the Thermodynamic Limit} 
\label{sec:thermodynamics_1}
For the Kuramoto-Sakaguchi model in the thermodynamic limit of infinitely many oscillators, relationships between $r$, $\Omega$ as function of the coupling strength $K$ and the phase-frustration parameter $\lambda$ can be derived from the self-consistency relation (\ref{self_consistency_real})--(\ref{self_consistency_imag}). For a Lorentzian intrinsic frequency distribution (\ref{lorentzian_distribution}), however, the Ott-Antonsen ansatz \cite{OttAntonsen08} provides a simpler way to derive the relationship. %, which is presented in Sec.~\ref{sec:thermodynamic_lorentzian}. 
%For a uniform intrinsic frequency distribution (\ref{uniform_distribution}), the integrals appearing in the self-consistency relation reduce to a set two algebraic equations, which can then be solved numerically. This is presented in Sec.~\ref{sec:thermodynamic_uniform}.
%%%%%%%%%%%%%%%%%%%%%%%%%%%%%%%%%%%%%%%%%%%
%\subsection{Lorentzian Frequency Distribution} 
%\label{sec:thermodynamic_lorentzian}
In the thermodynamic limit, a frequency-dependent version of the Ott-Antonsen ansatz was developped for the Kuramoto-Sakaguchi model (\ref{KS-1}) that is applicable for general intrinsic frequency distributions \cite{Omelchenko2012,Omelchenko2013} and also to a Kuramoto-Sakaguchi model with two populations of phase oscillators with different inter- and intra-population coupling strength in the context of chimera states \cite{AbramsEtAl_2008}. We apply an extended Ott-Anstonsen-ansatz following \cite{OttAntonsen08,AbramsEtAl_2008,Omelchenko2012} here to a Lorentzian frequency distribution (\ref{lorentzian_distribution}) to obtain explicit expressions for $r$ and $\Omega$ as functions of $K$ and $\lambda$.\\
%The Ott-Antonsen ansatz has been previously applied to a Kuramoto-Sakaguchi model with two populations of phase oscillators with different inter- and intra-population coupling strength in \cite{AbramsEtAl_2008} in the context of chimera states. Here we apply the OA-ansatz for the Kuramoto-Sakaguchi model (\ref{KS-1}) with one population of oscillators with a Lorentzian intrinsic frequency distribution (\ref{lorentzian_distribution}) following {OttAntonsen, O'melchenko, Abrams} , to obtain explicit expressions for r and \Omega as functions of K and \lambda

Recalling from Sec.~\ref{sec:mean_field}, in the thermodynamic limit, the phases in the Kuramoto-Sakaguchi model (\ref{KS-1}) are described by a normalized probability density function $\rho(\phi,t;\omega)$ satisfying the continuity equation (\ref{continuity_eqn_non_entrained}) (modulo a shift of the mean frequency). Following the Ott-Antonsen ansatz \cite{OttAntonsen08,Omelchenko2013}, the probability density function $\rho$ can be expressed in the form
\begin{equation}{\label{Ott_Antonsen_ansatz}} 
\rho(\phi,t;\omega)=\frac{1}{2\pi}\left\{1+\sum_{n=1}^{\infty}\left[\bar{z}^n(t;\omega)e^{n i \phi}+z^n(t;\omega)e^{-ni\phi}\right]\right\}.
\end{equation}
The ansatz (\ref{Ott_Antonsen_ansatz}) satisfies the continuity equation (\ref{continuity_eqn_non_entrained}) if $z$ lies on the so-called Ott-Antonsen manifold
\begin{equation} {\label{Ott_Antonsen_reduced_eqn}}
\dot{z}-i\omega z+\frac{1}{2}(Kre^{-i\psi}z^2e^{i \lambda}-Kre^{i\psi}e^{-i\lambda})=0
\end{equation}
with
\begin{equation} {\label{mean_field_1}}
re^{i\psi}=\int_{-\infty}^{\infty}z(t;\omega)g(\omega)d\omega=:\mathcal{G}z.
\end{equation}
The function $z(t;\omega)$ can be analytically extended to the upper half of the complex-$\omega$ plane, and $|z(t;\omega)|\rightarrow 0$ as $\text{Im}(\omega) \rightarrow \infty$. For a Lorentzian intrinsic frequency distribution (\ref{lorentzian_distribution}), which we recall here, 
\begin{equation*}
g(\omega)=\frac{\Delta}{\pi(\Delta^2+\omega^2)}=\frac{1}{2\pi i}\left(\frac{1}{\omega-i\Delta}-\frac{1}{\omega+i\Delta}\right),
\end{equation*}
the integral $\mathcal{G}z=\int_{-\infty}^{\infty}z(t;\omega)g(\omega)d\omega$ can be computed by completing a contour in the upper-half $\omega$ plane and applying the Residual Theorem, which yields $\mathcal{G}z=z(t;i\Delta)$. Then from (\ref{mean_field_1}) we obtain
\begin{equation} {\label{OA_mean_field_2}} r(t)e^{i\psi(t)}=\mathcal{G}z=z(t;i\Delta), 
\end{equation}
i.e. the value of the order parameter depends on the value of the function $z=z(t;\omega)$ at $\omega=i\Delta$ only.
In (\ref{Ott_Antonsen_reduced_eqn}), setting $\omega=i\Delta$ and substituting (\ref{OA_mean_field_2}) gives
\begin{align} 
\dot{r}&=-\frac{1}{2}Kr\cos\lambda\left(r^2-1+\frac{2\Delta}{K\cos\lambda}\right)\label{OA_reduced_equation_lorentzian_1}\\
r\dot{\psi}&=-\frac{1}{2}Kr\sin\lambda(1+r^2).
\label{OA_reduced_equation_lorentzian_2}
\end{align}
For $0\leq\lambda<\frac{\pi}{2}$, $r=0$ is a stable stationary solution of (\ref{OA_reduced_equation_lorentzian_1}) for $0<K<K_c$, where
\begin{equation} {\label{Kc_OA_lorentzian}}
K_c=\frac{2\Delta}{\cos\lambda}.
\end{equation}
For $K\geq K_c$, a pair of stable stationary solutions of (\ref{OA_reduced_equation_lorentzian_1})
\begin{equation} \label{r_OA_lorentzian}
r=\pm\sqrt{1-\frac{2\Delta}{K\cos\lambda}}=\pm\sqrt{1-\frac{K_c}{K}}
\end{equation}
emerges via a supercritical pitchfork bifurcation at $K=K_c$. The positive solution corresponds to the partially synchronized state, and thus $K_c$ marks the  onset of partial synchronization. Substituting the positive solution of $r$ into (\ref{OA_reduced_equation_lorentzian_2}) yields the cluster mean frequency 
\begin{equation} \label{Omega_OA_lorentzian}
\Omega=\dot{\psi}=\Delta\tan\lambda-K\sin\lambda.
\end{equation}

In Sec.~\ref{sec:results}, the order parameter $r=r(K,\lambda)$ (\ref{r_OA_lorentzian}) and the cluster mean frequency $\Omega=\Omega(K,\lambda)$ (\ref{Omega_OA_lorentzian}) are compared with the corresponding values obtained from numerical simulations of the full Kuramoto-Sakaguchi model (\ref{KS-1}) (cf. Fig.~\ref{fig:comparison_r}(a) and Fig.~\ref{fig:comparison}(a,c)); numerical results for the critical coupling strength $K_c=K_c(\lambda)$ (\ref{Kc_OA_lorentzian}) are presented in Fig.~\ref{fig:kc_comparison}(a).

%%%%%%%%%%%%%%%%%%%%%%%%%%%%%%%%%%%%%%%%%%%

%%%%%%%%%%%%%%%%%%%%%%%%%%%%%%%%%%%%%%%%%%%

\bibliographystyle{apsrev4-1}
%\bibliography{KuramotoSakaguchi}{}

%merlin.mbs apsrev4-1.bst 2010-07-25 4.21a (PWD, AO, DPC) hacked
%Control: key (0)
%Control: author (72) initials jnrlst
%Control: editor formatted (1) identically to author
%Control: production of article title (-1) disabled
%Control: page (0) single
%Control: year (1) truncated
%Control: production of eprint (0) enabled
%

\end{document}